\documentclass{aa}
\usepackage{natbib}
\usepackage{xspace}
\usepackage{amssymb}
\usepackage{amsmath}
\usepackage{hyperref}
\usepackage{graphicx}
\usepackage{lineno}

\def\hatG{{\hat G}}
\def\barG{{\tilde G}}

\def\stoiphase{fixed-composition phase}

\def\phasehull{\textsc{PhaseHull}\xspace}

\begin{document}
\setlength\linenumbersep{3pt}
\title{Computing phase diagrams using a convex hull algorithm}
\titlerunning{Computing phase diagrams using a convex hull algorithm}
\authorrunning{Dullemond \& Young}
\author{C.~P.~Dullemond$^{1}$ and E.~D.~Young$^{2}$}
\institute{(1) Institute for Theoretical Astrophysics (ITA), Center
  for Astronomie (ZAH), Heidelberg University, Albert Ueberle Str.\ 2,
  69120 Heidelberg, Germany; Email: dullemond@uni-heidelberg.de\\
(2) Department of Earth, Planetary, and Space Sciences, University of California, Los Angeles, CA 90095, USA, Email: eyoung@epss.ucla.edu
}
\date{\today}

\abstract{
We present a simple universal computational algorithm for computing {compositional} phase diagrams of rocks and their melts {at given temperature and pressure}. It makes use of the mathematical concept of the convex hull of a set of points in the space spanned by the composition and the Gibbs free energy. All the complexities of determining the stability or separation of phases, the localization and orientation of tie lines, as well as the determination of characteristic points, curves and surfaces such as the solidus, liquidus, solvus, and the eutectic/peritectic points etc, are taken care of by the algorithm that computes the convex hull, supplemented with an algorithm to physically classify the resulting simplices. For the convex hull computation, the publicly available Qhull package can be used, which is available in SciPy. This makes this method accessible and intuitive for a broad set of {scientific and educational} applications. {Although the method is not practical for systems of a large number of components, it is remarkably stable and efficient for systems of up to four.} We present our implementation of the method as a publicly available Python package.
}

\maketitle


\section{Introduction}
Phase diagrams for solids, liquids and gases play a key role in many fields of science and engineering. In astrophysics and geophysics they are used, for instance, to determine the state and composition of cosmic dust particles, meteorites, and planetary interiors. These diagrams can be computed from the molar Gibbs free energy $\hatG(T,P;x_0,\cdots,x_{M-1})$ as a function of temperature $T$, pressure $P$, and the relative mole fractions ${\mathbf x}=(x_0,\cdots,x_{M-1})$ of the $M\ge 1$ system components, {where $\sum_0^{M-1}x_i=1$}. As an example, silica (SiO$_2$), magnesia (MgO) and lime (CaO) constitute a system of $M=3$ components that define solid mineral phases such as forsterite (Mg$_2$SiO$_4$), enstatite (MgSiO$_3$), and diopside (CaMgSi$_2$O$_6$), to name a few. Enstatite and diopside can either be distinct stoichiometric phases, or the endmembers of a continuous solid solution series in the pyroxene phase.

Depending on temperature and pressure, some minerals become unstable and disappear, while others become stable and emerge. At elevated temperatures, mixtures of solids and liquids can form, and at even higher temperatures, all condensed material melts to form a liquid. A phase diagram is a visual tool that indicates, for a given temperature and pressure, which combinations of coexisting stable phases emerge as a function of bulk composition ${\mathbf x}$ at chemical equilibrium.

Computing a phase diagram amounts to finding the linear combination of coexisting phases that has the lowest Gibbs free energy. The Gibbs free energy is the Legendre transform of internal energy that applies where the independent variables are pressure, temperature, and composition (rather than volume, entropy, and composition, for example). This choice of energy is particularly well suitable for studying solids and liquids because of their near-incompressibility, and because many systems in astro- and geophysics are close to hydrostatic equilibrium.

A commonly used way of finding the co-existing stable phases is to apply, for each bulk composition ${\mathbf x}$, temperature $T$ and pressure $P$ of interest, a minimization solver from a standard numerical library. In Appendix \ref{sec-app-gibbs-minimization} we will discuss an implementation of such a method. This method gives the phase composition (the coexisting phases and their relative abundances) at a specific point in the phase diagram, and is therefore well suited for use in simulation software, or when studying a material with a known bulk composition. However, computing a complete {compositional} phase diagram with this approach would require applying the minimization solver to a grid of values of $(x_0,\cdots,x_{M-1})$, making it rather inefficient.

Alternatively, and complementary to this, one can identify, at a given $T$ and $P$, the locations of  tie lines connecting compositions of phases at equilibrium, the binodal curves comprising boundaries between miscibility and immiscibility, and the liquidus and solidus curves that define compositions of coexisting liquids and solids in equilibrium, etc within the $(x_0,\cdots,x_{M-1})$ component space. This approach gives the full {compositional} phase diagram at once, and gives more insight into why certain phases are stable and others are not. However, its computation can be cumbersome and often requires a combination of different methods, calculations and checks.

We present a convenient and underutilized, if not entirely ``new'', computational approach to computing phase diagrams, emphasizing the application to relevant binary, {ternary and quaternary} systems for easy visualization. It employs the mathematically well-defined concept of the {convex hull} or {convex envelope} of a finite or infinite set of points in an $M$-dimensional space. These points are constructed from the $M-1$ independent mole fractions, and their corresponding Gibbs free energies, together making up an $M$-dimensional composition-Gibbs space. Any point lying on the bottom (low-Gibbs free energy) side of the convex hull spanned by these points is stable, while any point above it is unstable. By definition none of the points lies below the convex hull. The construction of the convex hull of these points yields the complete set of stable phases, as well as the tie lines, tie triangles, tie tetrads, binodal curves (e.g., solvi), and liquidi and solidi, defining regions of coexisting phases and their compositions. These features comprise the desired phase diagrams.

The basic idea of the approach is not new at all: It is well known in material sciences that the stable phases all lie at the bottom of the convex hull of all conceivable phases comprising the Gibbs free energy surface at a given pressure and temperature  \citep[e.g.,][]{hautier:2014, brentfultz:2000, 2000PhRvB..62.3648W, 2008JMatR..23.1398G}. Some computer codes already use this concept for their internal workings \citep[e.g.][]{2010RJPCA..84..525V, ONG2013314, aflowchull_oses_2018}. In this paper, we introduce this concept to the astrophysics and geophysics community, for computations of stable solids and melts.

The method we present is to compute this convex hull, to isolate its bottom part (the downward facing facets), and to classify the physical meaning of each of these facets. Mathematically these facets are simplices, i.e., in 1D (a binary system) they are line elements, in 2D (a ternary system) they are triangles, in 3D (a quaternary system) they are tetrads, and so on for higher dimensions. The end-product is a set of simplices covering the phase diagram, with each simplex connecting stable phases at its corner points. This set of simplices, if computed from a complete (or at least sufficiently densely sampled) set of compositional points, is guaranteed to represent the unique and correct lowest-Gibbs free energy configuration. The only remaining challenge is then to correctly identify the physical meaning of each of these simplices.

One complication is that, once a variable-composition phase (a liquid or solid solution) is involved, the convex hull involving a continuous function must be determined. This can be done by sampling the function on a discrete grid of composition points $\{{\mathbf x}_j\}$ at the pressure and temperature of interest {\citep{2005E&PSL.236..524C}}, but leads to a large number of points in $({\mathbf x}_j,G_j)$-space for which the convex hull needs to be determined.

Fortunately, the problem of determining the convex hull of a large set of points in a higher-dimensional space is a solved mathematical problem. There exist publicly available, free libraries that solve it efficiently. This means that with our algorithm, all the ``hard work'' is delegated to this external mathematical library, leaving us only with the task of classifying the resulting simplices to give them physical meaning.

In addition to describing the algorithm, we also present the computer program called \phasehull{}, a Python package for employing the convex hull method to compute phase diagrams in geophysics and astrophyics. It uses automatic grid refinement near the liquidi and binodal lines. With \phasehull{} one can also compute the activities of the components in the liquid part of the material, in regions of coexisting solids and liquids. This is required for computations of evaporation and condensation of rocky materials at high temperatures, for example, a common problem in astrophysical and planetary science applications.

\phasehull{} is open source and can be downloaded from github. We tested the method on a variety of problems for which the correct solutions are well described in the literature. We also tested the method on particularly challenging cases with nearly-but-not-entirely flat regions in their Gibbs free energy surfaces. We find remarkable stability and reliability. If the method fails, it is usually either a resolution problem, or a misclassification of simplices in the correct convex hull, both problems that can be fixed with minor fine-tuning of the algorithm.

\phasehull{} {cannot, and is not meant to,} replace widely used software {such as, e.g., MELTS, Perple\_X and many others}. In particular, it is not plug-and-play software that can be used in a web or GUI interface or excel table. {And it is limited in the number of system components it can handle efficiently.} {Also, its primary strength lies in compositional phase diagrams at given temperature $T$ and pressure $P$, while it performs less well compared to other software for $T-P$ phase diagrams computed at a single given composition ${\bf x}$.} But it may serve as a useful complementary method, allowing experimentation with new thermodynamic data, serving educational purposes, and/or providing a relatively simple tool to be implemented in a variety of modeling frameworks.

The paper is structured as follows. First, in Section \ref{sec-method}, we will outline the convex hull method. Then, in Section \ref{sec-examples}, we will present a series of examples, and while doing so, we will explain more details of the method. We will then discuss advantages and disadvantages compared to standard approaches and conclude. The appendices contain many of the technical details. Also in the appendix is a description of a complementary (and to most readers perhaps more familiar) method of Gibbs minimization. This serves to put the convex hull method into context, and assures that the \phasehull{} software package is flexible, in the sense of offering both options to the user.

\section{Convex hull method}
\label{sec-method}

\subsection{Basic principle}
\begin{figure}
  \includegraphics[width=0.95\linewidth]{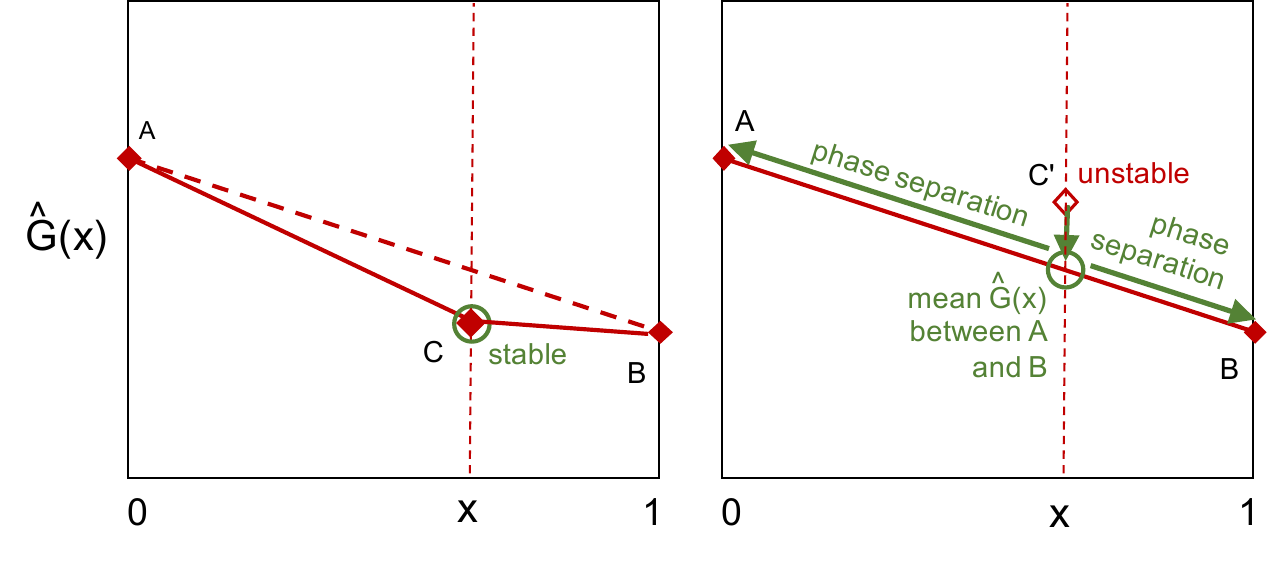}
  \caption{\label{fig-cartoon-phase-separation}Cartoon illustrating the stability/instability of a substance C consisting of $(1-x)A+xB$ and Gibbs free energy $\hatG(x)$ in a binary system consisting of components A and B. Left: Phase C is stable at the composition specified by $x$. Right: Phase C is unstable and separates into phases with compositions A ($x = 0$) and B ($x = 1$). Green circle is the final state Gibbs free energy.}
\end{figure}

The fundamental principle underlying the use of a convex hull algorithm for computing phase diagrams relies on the following thermodynamic rules:
\begin{enumerate}
\item Any set of phases $A$, $B$, etc can be {linearly} combined by keeping them unmixed (i.e., spatially separated on the microscopic scale). The combined molar Gibbs free energy of these coexisting phases is then:
\begin{equation}\label{eq-gibbs-linear-combination}
  \hatG_{\mathrm{total}} = \epsilon_A\hatG_{A} + \epsilon_B\hatG_{B} + \cdots
\end{equation}
where $\epsilon_{A,B,\cdots}$ are the mole fractions of the phases obeying $\sum_{k=A,B,\cdots}\epsilon_k=1$, and $\hatG_A$, $\hatG_B,\cdots$ are the molar Gibbs free energies {(to be defined more precisely in Section \ref{sec-scaled-formula-units})} of each of the phases relative to some suitable reference (e.g., free energies of formation from the elements or oxides). An example of such coexisting phases is the presence in a rock of distinct olivine and pyroxene minerals.
\item Sometimes it is energetically more favorable for these phases $A$, $B$, etc to mix at the atomic level, leading to a {solution}. An example of a solid solution is enstatite and diopside in a pyroxene mineral. Melting of rock typically leads to a well-mixed liquid solution (magma). In these cases, Eq.~(\ref{eq-gibbs-linear-combination}) will acquire additional non-linear terms due to entropy of mixing and possible interaction terms. In the context of the convex hull algorithm, such solutions will be precomputed, and treated as continuous series of phases that can be linearly combined with other phases using Eq.~(\ref{eq-gibbs-linear-combination}). In other words: The convex hull algorithm will only have to deal with the linear Eq.~(\ref{eq-gibbs-linear-combination}), even if non-linear mixing of phase components are involved. 
\item A phase is unstable if one can find a linear combination of other phases (together having the same bulk composition) that has a lower combined Gibbs free energy $\hatG_{\mathrm{total}}$ than that of the original phase $\hatG_{\mathrm{orig}}$ (see Fig.~\ref{fig-cartoon-phase-separation}-right). The original phase will then separate into the linear combination of phases that has the lowest combined Gibbs free energy $\hatG_{\mathrm{total}}$.
\item The relative fractions, by mole, of these emerging coexisting phases with compositions $A$ and $B$ can be computed using the lever rule (see appendix \ref{sec-app-lever-rule}). The lever rule together with Eq.~(\ref{eq-gibbs-linear-combination}) constitutes the linear interpolation of the Gibbs free energy between the  emerging phases, at the composition of the original unstable phase.
\end{enumerate}
The task of computing the phase diagram is to find, for any given bulk composition ${\bf x}$, the combination of coexisting phases that yields the lowest combined Gibbs free energy. This is an entirely linear problem as it is simply a physical combination of non-mixing phases. The non-linearities occur only before applying the convex hull algorithm. Fig.~\ref{fig-cartoon-phase-separation} shows, for discrete (fixed-composition, non-mixing) phases, how the stability or instability of a phase works.

\subsection{System components and scaled formula units}
\label{sec-scaled-formula-units}
In this paper, every phase is expressed in terms of the chosen system components. For example, system components SiO$_2$, CaO, and MgO comprise a ternary system ($M = 3$) and can serve as the apices of a ternary phase diagram. Minerals such as Mg$_2$SiO$_4$ (forsterite), MgSiO$_3$ (enstatite), CaMgSi$_2$O$_6$ (diopside) and CaSiO$_3$ (wollastonite) can all be formulated as composed of these three system components. The mole fraction compositions ${\mathbf x}$ of these stoichiometric phases are $(1/3,0,2/3)$, $(1/2,0,1/2)$, $(1/2,1/4,1/4)$, and $(1/2,1/2,0)$, respectively.

At a given temperature $T$ and pressure $P$, each of the minerals (let us identify them using an index $k$) has a molar Gibbs free energy $\barG_k$, which is the Gibbs free energy per mole of formula unit of that mineral. The values of $\barG_k(T,P)$ can be computed using tabular data from the literature (see appendix \ref{sec-databases}). 

However, one mole of  system components produces, in this example, only $1/3$, $1/2$, $1/4$, and $1/2$ moles of formula unit of the respective minerals. In the case of the mineral forsterite: 1/3 mole of SiO$_2$ and 2/3 mole of MgO (together 1 mole of system component) produce 1/3 mole of Mg$_2$SiO$_4$. Let us define $s$ as the scale factor, such that one mole of  system components produces $1/s$ moles of the mineral. To be able to embed the minerals into the ternary phase diagram, we must scale the formula units of the minerals by $1/s$. The scaled formula units of the above minerals then become Mg$_{2/3}$Si$_{1/3}$O$_{4/3}$, Mg$_{1/2}$Si$_{1/2}$O$_{3/2}$, Ca$_{1/4}$Mg$_{1/4}$Si$_{1/2}$O$_{3/2}$, and Ca$_{1/2}$Si$_{1/2}$O$_{3/2}$, respectively. Their scaled molar Gibbs free energies become
\begin{equation}\label{eq-scaling-gibbs}
\hatG_k = \frac{\barG_k}{s_k}
\end{equation}
where $s_k$ is the $s$ scaling factor of each mineral (for forsterite this would be $s_k=3$). We will always use these scaled versions of the minerals, except when mixing them in solid solutions (see Section \ref{sec-solid-solutions}), in which case it matters which multiplicity of the formula unit is used to compute the entropy of mixing. But even then, once the proper mixing entropy is computed, everything is scaled again.

In this way, all extensive quantities of the phases are defined ``per mole of  system components''. 

Note that the components also constitute their own phases. In the current example, they would be quartz (or one of its polymorphs) for SiO$_2$, lime for CaO, and periclase for MgO.

Once the phase diagram has been computed, the mole fractions $Y_k$ of each of the coexisting phases, for any given bulk composition ${\mathbf x}$, can be reconstructed using the lever rule (see appendix \ref{sec-app-lever-rule}). The $Y_k$ refer to the moles of scaled formula units per mole of  system components. In the above example, $Y_{\mathrm{Mg}(2/3)\mathrm{Si}(1/3)\mathrm{O}(4/3)}$ would refer to the amount of moles of the mineral Mg$_{2/3}$Si$_{1/3}$O$_{4/3}$ present, when starting from one mole, in total, of SiO$_2$ and MgO. The amount of moles of un-scaled formula unit (here: Mg$_2$SiO$_4$) is then easily obtained as $Y_k/s_k$.

\subsection{Fixed-composition phases}
\label{sec-fixed-composition-phases}

The simplest phases we consider are fixed-composition phases. Typically these are stoichiometric solid minerals such as the examples of Mg$_2$SiO$_4$, MgSiO$_3$, CaMgSi$_2$O$_6$ and CaSiO$_3$ mentioned above. In the phase diagram they are represented by points with well-defined coordinates $(x_0,\cdots,x_{M-1})$, like those depicted in Fig.~\ref{fig-cartoon-phase-separation}. 

In the real world they are likely not {exact} points, in the sense that small deviations from the precise stoichiometry can occur. Each fixed-composition phase represents the lowest point of a very narrow and steep Gibbs free energy surface. The narrower this Gibbs free energy profile, the tighter the composition is defined. Fixed-composition phases are the approximation that this Gibbs free energy surface is an infinitely narrow ``needle'' pointing downward, the lowest point being the Gibbs free energy of the phase.

For the convex hull algorithm, fixed-composition phases are the cheapest to handle, as each phase is represented by only a single point in the phase diagram. These phases are, by definition, assumed to be unmixed.

\subsection{Liquids}
\label{sec-liquid-solutions}

If a mineral substance melts, it becomes a liquid solution (a magma). In such a liquid, all system components are assumed to be mixed at the atomic level. The molar Gibbs free energy of the liquid, at any given bulk composition ${\mathbf x}$, is
\begin{equation}\label{eq-gibbs-liquid-solution}
\hatG_{\mathrm{liq}}({\mathbf x}) = \sum_i x_i\mu^{\mathrm{liq},0}_{i} + RT \sum_i x_i\ln(x_i) + \hatG_{\mathrm{liq}}^{\mathrm{ex}}({\mathbf x})
\end{equation}
where $\mu^{\mathrm{liq},0}_{i}$ are the Gibbs free energies of the liquid phase components, and $\hatG_{\mathrm{liq}}^{\mathrm{ex}}$ is the excess Gibbs free energy caused by the interatomic forces between the mixed components. In this equation, the first term on the right-hand-side is the linear combination term, the second term is the contribution of the entropy of mixing, and the final term (the interaction term) describes the deviation from ideal mixing. Magmas are highly non-ideal solutions, meaning that the final term in Eq.~(\ref{eq-gibbs-liquid-solution}) is comparable to, or larger than, the entropy of mixing term\footnote{Note that other ways exist to describe magmas, in which the interaction term is strongly reduced \citep{1980CoMP...71..323G} or even vanishes \citep{1987E&PSL..82..207F}.}. The key to a reliable description of the magmatic liquid (or any solution for that matter) is a good model for the excess Gibbs free energy $\hatG_{\mathrm{liq}}^{\mathrm{ex}}({\mathbf x})$ (see appendix \ref{sec-precomputing-Gibbs}).

A gas (vapor) phase can be treated in the same way as a liquid. Non-ideality can emerge, encoded by fugacity coefficients, which correct for both non-ideal mixing and non-ideal PVT behaviors \citep{2026ApJ...999..178W}.

\subsection{Solid solutions}
\label{sec-solid-solutions}

Like liquids, also solids can mix at the atomic level, as we mentioned before in the example of the enstatite-diopside join. This is called a solid solution. However, many solid phases, in particular among the oxides, are immiscible, for instance due to incompatible crystal structures. The most extreme cases would be the fixed-composition phases mentioned in Section \ref{sec-fixed-composition-phases}. A particularly interesting intermediate example, involving both miscibility and immiscibility, is the existence of two mutually-immiscible solid solution phases in the SiO$_2$, FeO, and MgO ternary system ($M = 3$). The two phases are the olivine phase, being a solid solution between the phase components Mg$_2$SiO$_4$ (forsterite) and Fe$_2$SiO$_4$ (fayalite), and the pyroxene phase, being a solid solution between the phase components MgSiO$_3$ (enstatite) and FeSiO$_3$ (ferrosilite). In this case there are three system components that describe the compositions of four phase components that in turn apply to two distinct phases (olivine and pyroxene) that are each continuous solid-solution mixtures, but are not miscible with each other.  We might express the molar Gibbs free energy of olivine $\hatG_{\rm oli}$ in terms of the mole fractions of Mg$_2$SiO$_4$ and Fe$_2$SiO$_4$,
\begin{equation}\label{eq-gibbs-solid-solution}
\begin{split}
  \hatG_{\rm oli}(y_{\rm fo},y_{\rm fa}) =& y_{\rm fo}\mu^0_{\rm fo} + y_{\rm fa}\mu^0_{\rm fa} \\
  & +RT y_{\rm fo}\ln(y_{\rm fo}) +RT y_{\rm fa}\ln(y_{\rm fa}) \\
  &+ \hatG_{\mathrm{oli}}^{\mathrm{ex}}(y_{\rm fo},y_{\rm fa})
\end{split}
\end{equation}
where $y_{\rm fo}$ and $y_{\rm fa}$ sum to unity\footnote{We use $y$ instead of $x$ to distinguish the {phase components} forsterite and enstatite from the {system components} silica, ferrous oxide and magnesia.}.  The mixtures of forsterite and fayalite can in turn be described as mole fractions of the system components for the purposes of applying the convex hull algorithm. This is because Mg$_2$SiO$_4$ can be described as being composed of 2MgO + SiO$_2$, and therefore, in terms of mole fractions of system components, it is described as $x_{\rm MgO} = 2/3$, $x_{\rm SiO_2} = 1/3$, and  $x_{\rm FeO} = 0$, and similarly for the iron endmember.  Since there are three system components and four phase components, there should be one linearly independent reaction among these phases.  We can write this reaction as 2MgSiO$_3$ + Fe$_2$SiO$_4$ $\rightleftharpoons$ 2FeSiO$_3$ + Mg$_2$SiO$_4$.   A tie line connecting the compositions of these two phases where both are stable in the ternary phase diagram depicts the equilibrium state described by this reaction.  

Note that if we choose the phase components of a solid solution as the system components of our phase diagram (here they would then be binary phase diagrams), then we can treat a solid solution with the same machinery as for a liquid, and $y$ becomes $x$. In general, though, solid solutions form subspaces (hypersurfaces) in the full phase space of the system. It is for this reason alone that solid solutions form a special class of phases in the \phasehull{} software package, distinct from the liquid class.

\subsection{Creating a point cloud in composition-Gibbs space}
\label{sec-creating-points}

The strategy for constructing the phase diagram, in the algorithm presented here, is to pre-compute, for a given temperature $T$ and pressure $P$, the Gibbs free energies $\hatG_k$ of all possible phases ${\bf x}_k$, whether they are stable or not. The continuum phases (liquid and solid solutions) require a grid of compositional sampling points ${\bf x}_{k,n}$ with $0\le n<N_k$, where $N_k$ is the number of grid points for phase $k$. Each point $k,n$ has its own pre-computed value of $\hatG_{k,n}=\hatG_k(x_{k,n})$ for that phase. For a liquid phase, the grid points ${\bf x}_{k,n}$ cover the full $M-1$-dimensional space spanned by the system components. The grid can be refined where necessary by inserting further points in between those of a base grid. For a solid solution phase, the gridpoints are confined to the (equal- or lower-dimensional) subspace spanned by its phase components, which can be different for different solid solutions. The phase diagram we wish to construct is the full $M-1$-dimensional space spanned by the system components.

Once all Gibbs free energies are computed for all phases (discrete and continuous), the reduced $x$-coordinates in terms of system components, $\mathbf{\bar x}_n=(x_0,\cdots, x_{M-2})$ (see appendix \ref{sec-app-lever-rule}), are combined with their Gibbs free energies $\hatG_k(\mathbf{x}_{k,n})$ into an $M$-dimensional vector
\begin{equation}
{\bf P}_{k(,n)}=(x_{k(,n),0},\cdots x_{k(,n),M-2},\hatG_{k(,n)})
\end{equation}
This set of points ${\bf P}_{k(,n)}$ is then fed into the convex hull algorithm, to be discussed below. 

\subsection{Convex hull of a set of points in $M$-dimensional space}

\begin{figure}
  \includegraphics[width=0.95\linewidth]{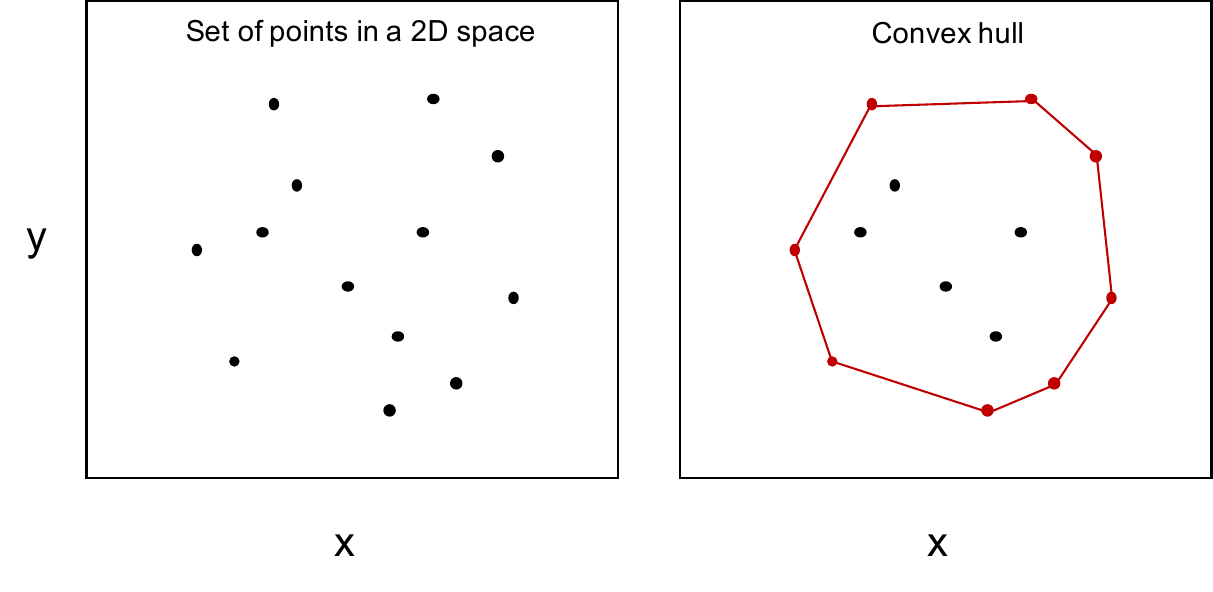}
  \caption{\label{fig-cartoon-convex-hull}Cartoon illustrating the convex hull of a set of points in 2D space. Right: The set of points. Left: The identification of the convex hull of that set of points. The red points are the vertices of the convex hull (the points lying on the convex hull), and the red lines are the simplices connecting the vertices. In N-dimensional space, these simplices will become triangles ($N=3$), tetrads ($N=4$), etc., and together they form a closed $N-1$-dimensional (hyper-)surface $\partial V$ enclosing the convex $N$-dimensional (hyper-)volume $V$ spanned by the set of points.}
\end{figure}

\begin{figure}
  \centerline{\includegraphics[width=0.80\linewidth]{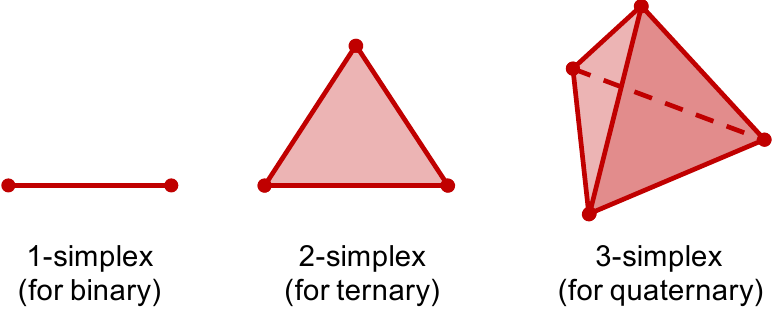}}
  \caption{\label{fig-cartoon-simplices}The basic building blocks of phase diagrams using the convex hull method: Simplices. For binary phase diagrams, they are line elements (left). For ternary phase diagrams, they are triangles (middle). For quaternary phase diagrams they are tetrads (right).}
\end{figure}

The convex hull of a set of points in an $M$-dimensional space is the tightest convex closed $M-1$-dimensional surface containing all these points. For $M=2$ this is illustrated in Fig.~\ref{fig-cartoon-convex-hull}. One can regard this as the shape of a rubber band that is stretched to contain all points, and then let go to tighten itself around these points. For $M=3$ the rubber band would be a rubber bag, but the principle remains the same. The convex hull consists of the list of points that lie on it, and the list of line elements (in 2D) or $M-1$-dimensional simplices (in $M$D), see Fig.~\ref{fig-cartoon-simplices}, that connect these points and together form the closed convex hull surface. 

Finding the convex hull for a large set of points in an $M$-dimensional space is not straightforward, but is a well-posed mathematical problem for which various libraries exist. The publicly available Qhull library\footnote{\url{http://www.qhull.org}} \citep{quickhull:1996, 2013ascl.soft04016B}, which is natively built into the SciPy library \citep{2020SciPy-NMeth} in Python, is perhaps the most accessible and reliable one. It returns the list of indices of the points that lie on the convex hull and a list of simplices of the convex hull. For each simplex, it lists the indices of the points that make up its corners, the indices of its neighboring simplices, and the outward-pointing normal vector and offset.

In the present case we have $M-1$ compositional dimensions ($\mathbf{\bar x}$) and one Gibbs free energy dimension $\hatG_{\rm total}$ for the system as a whole, which we shall assign the directional qualities ``up/above'' for higher values of $\hatG_{\rm total}$ and ``low/below'' for lower values of $\hatG_{\rm total}$.

The objective of finding the phase diagram is therefore to find the convex hull of the volume of all possible phases in this space, and isolate the bottom facets of this hull, because they represent the lowest energy (most favorable) combinations of phases. The bottom facets are easily identified by their normal vectors pointing downward, i.e., having a negative value in their $\hatG$-dimension.

\subsection{Application to phase diagrams}

To illustrate the application of the convex hull algorithm to the computation of phase diagrams, we present here a set of fictitious model binary systems.

\begin{figure}
  \includegraphics[width=\linewidth]{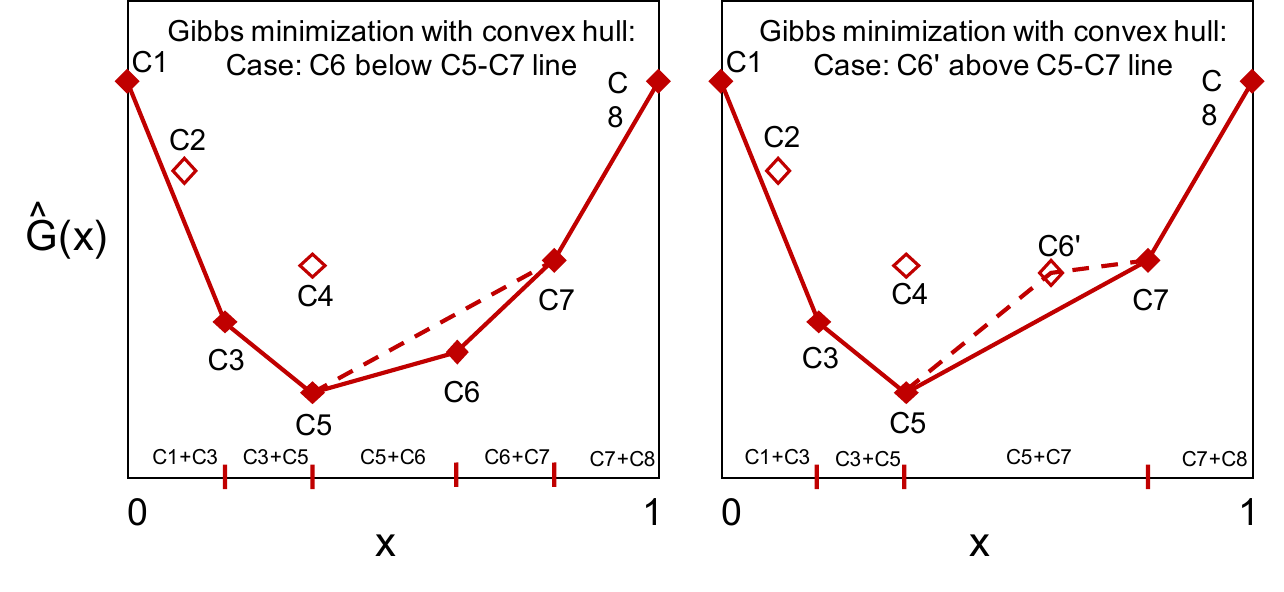}
  \caption{\label{fig-cartoon-crystals} Cartoon illustrating the convex hull method applied to a binary system consisting of a set of fixed-composition phases depicted with the red diamonds. Dashed lines are energetically unfavorable (forbidden) tie lines, solid lines are allowed tie lines. Left: case where \stoiphase{} C6 has a Gibbs free energy below the tie line between phases C5 and C7, in which case C6 belongs to the convex hull and splits the composition space between C5 and C7 into two regions with coexisting phases C5+C6 and C6+C7. Right: case where \stoiphase{} C6 (here denoted with an apostrophe) has a Gibbs free energy above the tie line between phases C5 and C7. In this case only C5 and C7 can coexist, and C6' is energetically unfavorable.}
\end{figure}

For a set of solid \stoiphase{}s with fixed composition in a binary system, the task of finding the phase diagram with the convex hull method is illustrated in Fig.~\ref{fig-cartoon-crystals}. The ``rubber band'' is tightened from below, and all phases that lie above it are unstable. The red lines between the stable \stoiphase{}s are tie lines that represent the coexistence between the phases on each side. A liquid melt with an arbitrary composition $x$ that is cooled down to low temperatures where all material is in a solid phase, will in general be forced to split into two adjacent \stoiphase{}s. The mole fractions of these two coexisting phases are determined by the lever rule. Only if the composition of the liquid happens to be exactly equal to one of the \stoiphase{}s, will the liquid and solid phase have the same composition. This is the case of ``congruent'' melting of the solid phase. For instance, at low pressures, Mg$_2$SiO$_4$ melts congruently.

\begin{figure}
  \includegraphics[width=\linewidth]{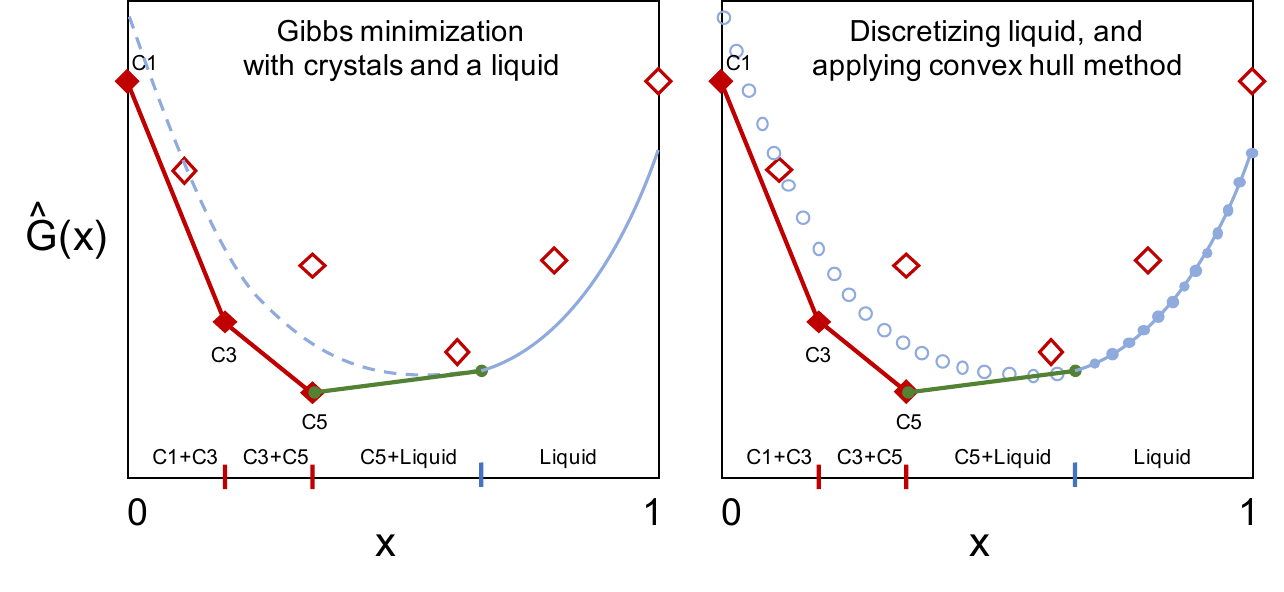}
  \caption{\label{fig-cartoon-crystliquid} Cartoon illustrating the convex hull method applied to a binary system consisting of a set of fixed-composition phases (red diamonds) and a liquid (blue line and symbols). Left: the true physical continuous case. Right: the discretized approximation used in the convex hull method. Open diamond symbols are unstable phases. Open circles are unstable liquid phase points. Red lines are tie lines between stable phases. Green line is the tie line between a fixed-composition solid phase and the liquid, where solid coexists with the liquid. This tie line is tangent to the liquid curve. Blue dots are the grid points and blue lines connecting them are the linear interpolations between them.}
\end{figure}

At high temperatures, solids generally melt and become liquid, which generally forms a continuous solution, as illustrated in Fig.~\ref{fig-cartoon-crystliquid}-left. By mapping this continuum on a grid, one obtains a discrete set of compositional points with their Gibbs free energy values, which, together with the discrete \stoiphase{}s, is fed into the convex hull algorithm, which then finds the stable phases and their coexistence tie lines, as illustrated in Fig.~\ref{fig-cartoon-crystliquid}-right. The liquidus, the $P$, $T$, and compositional state where a material first transitions from entirely liquid to liquid plus a solid, is the point where the green tie line is tangent to the liquid Gibbs curve. This illustrates the usual way of representing the equilibrium condition in a binary system; equilibrium is obtained where the stable phases share a common tangent.
This is equivalent to the statement that the chemical potentials on both ends of the tie line are equal (see appendix \ref{sec-app-liquidus} for a discussion).

The convex hull algorithm finds this point approximately, to the resolution set by the number of grid points. Obviously, the convex hull method gets more accurate in finding the liquidus location, the higher the number of grid points.  In the example given here solid phase C5 has the honor of being referred to as the ``liquidus'' phase, meaning the first (but not necessarily the only) solid to crystallize from the liquid. 

\begin{figure}
  \includegraphics[width=\linewidth]{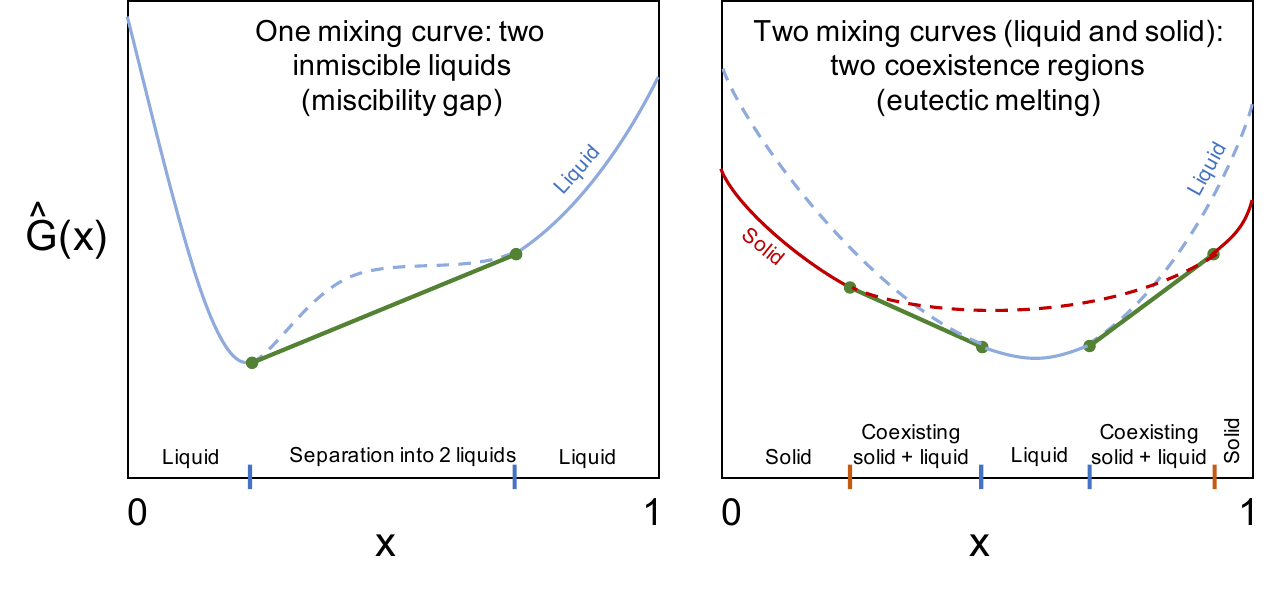}
  \caption{\label{fig-cartoon-continuous-phases} Cartoon illustrating the convex hull method applied to a binary system consisting of a liquid with a non-convex molar Gibbs free energy curve (left), and a binary system consisting of a liquid and a solid solution (right). In both cases the continua are discretized on a grid, as shown in Fig.~\ref{fig-cartoon-crystliquid}-right, and not repeated here. In both cases the non-convex part of the curve (left) or the combined curve $\mathrm{min}(\hatG_l,\hatG_s)$ (right) are bridged by tie lines. In the left case this is a region of immiscibility of the liquid. In the right case it is eutectic melting.}
\end{figure}

Tie lines can also emerge between two points of the same continuous phase, or between two different continuous phases, as illustrated in Fig.~\ref{fig-cartoon-continuous-phases}. In higher-dimensional systems (ternaries, quaternaries, etc), these tie lines can be challenging to compute with the classical method, but they emerge automatically with the convex hull method. 

\section{Examples and further details of the method}
\label{sec-examples}

To make it easier to present the details of the convex hull method, we break with convention by starting with example applications before fully completing the description of the computational method. We will outline the details as they emerge.

\subsection{The MgO -- SiO$_2$ binary system (Berman 1983)}
\label{sec-mgo-sio2}

Magnesia (MgO) and silica (SiO$_2$) form a well-studied binary system. It is part of the CaO-SiO$_2$-MgO-Al$_2$O$_3$ quaternary system model presented in the PhD thesis of \citet{BermanPhD1983}. We define $x=x_{\mathrm{MgO}}$ and $1-x=x_{\mathrm{SiO}_2}$. The excess Gibbs free energy $\hatG_{\mathrm{ex}}(x)$ of the liquid (melt) is given by fourth-order Margules parameters (see appendix \ref{sec-precomputing-Gibbs}), and especially for low temperatures the asymmetry becomes strong. All thermodynamic coefficients needed to construct this model in \phasehull{} are listed in \citet{BermanPhD1983}.

\begin{figure}
  \includegraphics[width=\columnwidth]{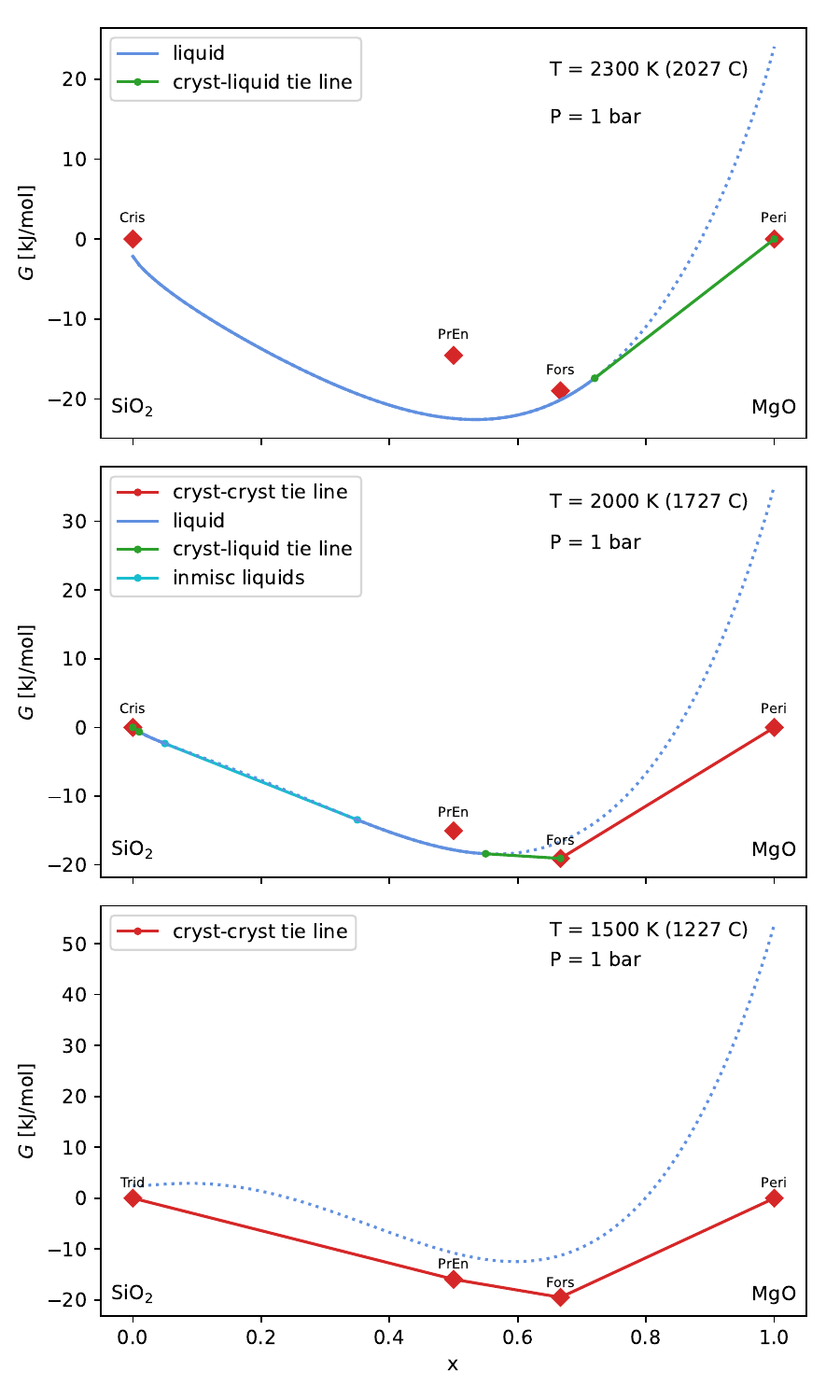}
  \caption{\label{fig-model-MgO-SiO2-x-G} Gibbs free energy as a function of composition for the MgO -- SiO$_2$ binary system, with the convex hull (i.e., the coexisting stable phases) shown. The baseline between the MgO and SiO$_2$ solid phases is subtracted, so that these system component solids are at $\hatG=0$, for clarity. Top to bottom: different temperatures. Dotted blue line: Gibbs free energy of the liquid phase, where the liquid phase is unstable. Solid blue line:  Gibbs free energy of the liquid phase, where the liquid phase is stable. Red diamonds: Fixed-composition phases (from left to right: quartz/tridymite, proto-enstatite, forsterite, periclase). Red lines: Tie lines connecting the stable solid phases. Green line(s): Tie line connecting a fixed-composition solid with the liquid. Cyan line (middle panel, between $x=0.05$ and $x=0.35$): immiscible liquids.}
\end{figure}

The model at three different temperatures is shown in Fig.~\ref{fig-model-MgO-SiO2-x-G}. At 1500 K (1227$^\circ$C) all phases are solid. The convex hull consists of only four points and three simplices (tie lines).

At 2000 K (1727$^\circ$C) both quartz/cristobalite and forsterite acquire a liquidus tie line (in green). Also an immiscibility region appears in the liquid (in cyan, {between $x=0.05$ and $x=0.35$}). The convex hull now consists of five distinct phase assemblages, depending on bulk compositions. Listed from low $x$ to high, they are:  1) cristobalite + melt; 2) two immiscible melt phases; 3) one melt phase; 4) melt + forsterite; and 5) forsterite + periclase. To the left of 1) is pure cristobalite, between 4) and 5) is pure forsterite, and to the right of 5) is pure periclase. The one melt phase of region 3) is represented by numerous `mini-tielines' between the gridpoints of the liquid (an artifact of the discretization).

At 2300 K (2027$^\circ$C) the only remaining solid phase is periclase, which is in equilibrium with a melt with a composition near to that of forsterite. Now the points are mostly the grid points of the liquid, plus the single point of periclase. The simplices are mostly the mini-tielines between the gridpoints of the liquid, and the single large tie line of the liquidus of periclase.

\begin{figure}
  \includegraphics[width=\columnwidth]{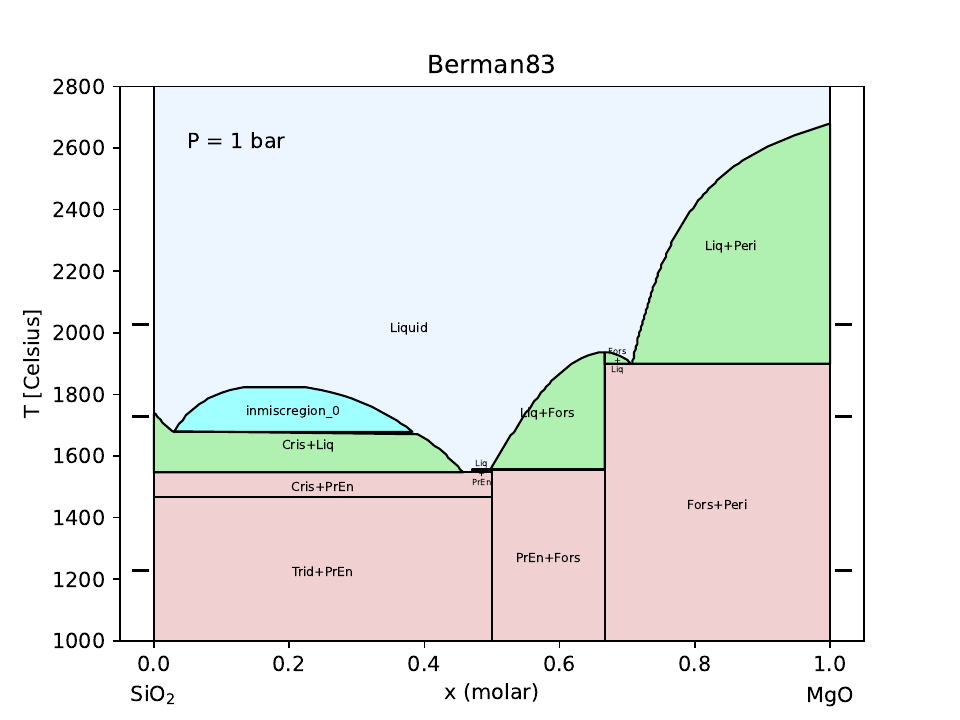}
  \caption{\label{fig-model-MgO-SiO2-x-T} Composition-temperature phase diagram of the same system as in Fig.~\ref{fig-model-MgO-SiO2-x-G}, {computed by applying the convex hull method for 200 discrete temperature values between 1000 and 2800 degrees Celsius}. Red: solid state coexistence regions of adjacent fixed-composition phases. Green: solid-liquid coexistence regions (tie lines between fixed-composition solid and its liquidus). Cyan: Immiscible liquids. Light blue: liquid. The three panels of Fig.~\ref{fig-model-MgO-SiO2-x-G} are horizontal cuts in this diagram at temperatures of 1227, 1727 and 2027 degrees Celsius (marked by dashes left and right of the axes in the figure). This diagram can be compared to Fig.\ 17 of Berman's PhD thesis, albeit with mole fraction on the $x$-axis rather than mass fraction.}
\end{figure}

By computing this phase diagram for a series of temperatures, one obtains the $x-T$ phase diagram, shown in Fig.~\ref{fig-model-MgO-SiO2-x-T}. It shows the liquidus (the curved lines between green and white), the solidus (the horizontal lines between green and red), and the region of immiscible liquids. 

\subsection{The CaO -- Al$_2$O$_3$ -- SiO$_2$ ternary system (Berman 1983)}
\label{sec-cao-al2o3-sio2}

The CaO-Al$_2$O$_3$-SiO$_2$ ternary system is a well studied mineral system with applications in industry (e.g., ceramics, glass, and cement). It is part of the CaO-SiO$_2$-MgO-Al$_2$O$_3$ quaternary system model presented in the PhD thesis of \citet{BermanPhD1983}, and the CaO-Al$_2$O$_3$-SiO$_2$ ternary system model of the paper of \citet{1984GeCoA..48..661B}. Both models are nearly, but not entirely, identical in their thermodynamic values. We use those of \citet{BermanPhD1983} here. There exist, however, also more detailed models of this system \citep[e.g.,][]{MaoHillert:2006, TanShi:2024}.

\begin{figure}
  \includegraphics[width=1.1\linewidth]{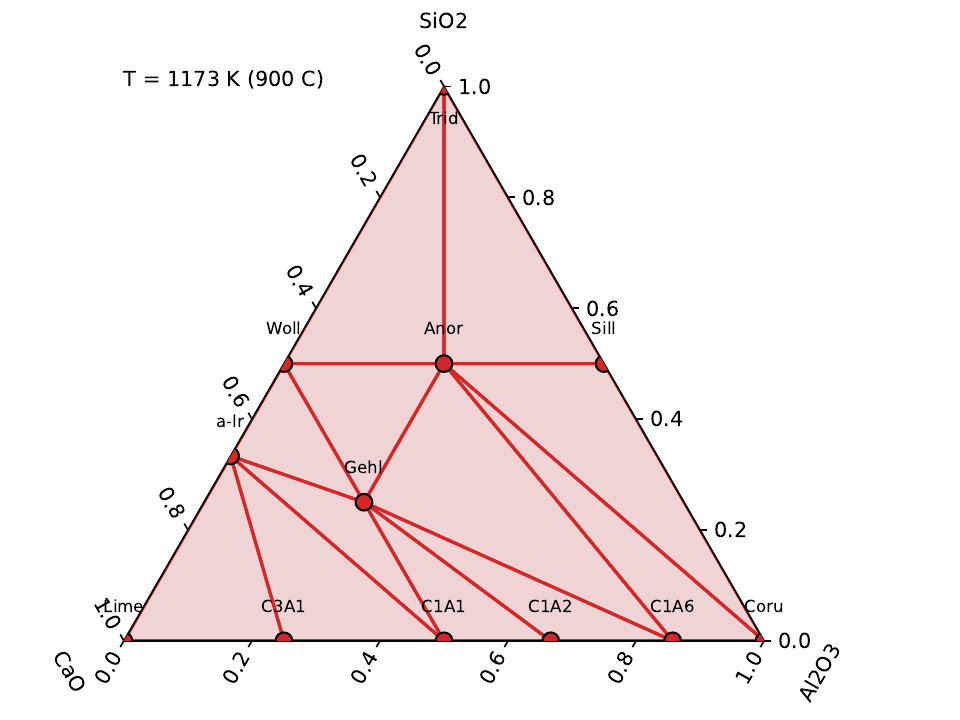}
  \caption{\label{fig-model-CaO-Al2O3-SiO2-lowtemp} Phase diagram of the CaO-Al$_2$O$_3$-SiO$_2$ ternary system of the \citet{BermanPhD1983} model at T=900$^\circ$C, as computed using the convex hull method, for a temperature of 900 Celsius, for which all stable phases are solid \stoiphase{}s. The convex hull algorithm provides the triangulation, i.e., it divides the surface of the phase diagram into simplices. Each simplex (triangle) is a region of coexistence of three solid \stoiphase{}s. For any point inside such a simplex, or on a tie line, the lever rule is used to determine the mole fraction of each of the three phases.}
\end{figure}

The simplest phase diagram is obtained at the (relatively) low temperature of 900$^\circ$C = $1173.15$K, as shown in Fig.~\ref{fig-model-CaO-Al2O3-SiO2-lowtemp}. At this temperature, the diagram contains only \stoiphase{}s, and no liquid phase is present. This is, in essence, the top-down view of the bottom of the convex hull. When comparing this ternary phase diagram to the cartoon binary case of Fig.~\ref{fig-cartoon-crystals}, our view would be from the top, looking down: where there is a red line in Fig.~\ref{fig-cartoon-crystals}, there is a red triangle (simplex) in Fig.~\ref{fig-model-CaO-Al2O3-SiO2-lowtemp}, and where there is a red diamond in Fig.~\ref{fig-cartoon-crystals}, there is a red circle in Fig.~\ref{fig-model-CaO-Al2O3-SiO2-lowtemp}. The red lines in the ternary diagram are the tie lines connecting pairs of \stoiphase{}s. The red simplices connect triples of such \stoiphase{}s. The fact that a simplex in 2D has, at most, 3 corners shows that at most three phases can co-exist, which is an illustration of Gibb's phase rule. 

\begin{figure}
  \includegraphics[width=1.1\linewidth]{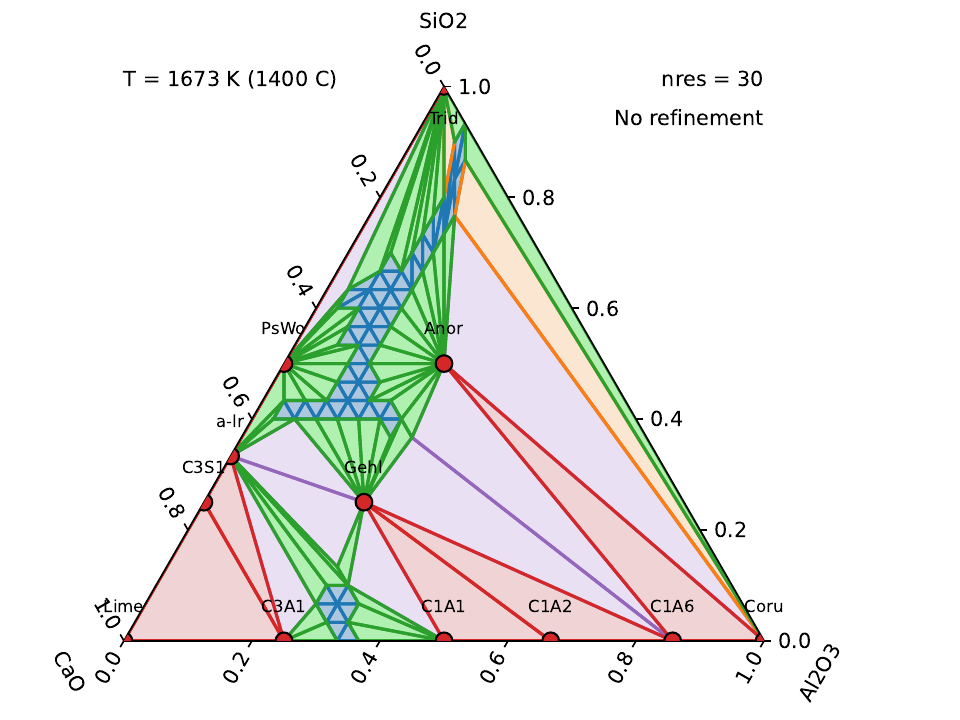}
  \caption{\label{fig-model-CaO-Al2O3-SiO2-lowres} Low-resolution phase diagram of the CaO-Al$_2$O$_3$-SiO$_2$ ternary system of the \citet{BermanPhD1983} model at T=1400$^\circ$C, as computed using the convex hull method, with a base resolution of $N_0=30$ and no grid refinement. The convex hull algorithm provides the triangulation. The physical meaning of each triangle (indicated with color) has to be identified in post-processing. See text for simplex classification strategies. The colors are as follows: Red circles are solid \stoiphase{}s, red triangles are three-phase coexistence simplices, lilac triangles are two-solid-one-liquid coexistence simplices, green triangles are 2D approximations of 1D tie lines, blue triangles represent the liquid.}
\end{figure}

If we now increase the temperature to 1400$^\circ$C = $1673.15$K, a liquid phase emerges in part of the phase diagram, introducing the need for a grid to map the continuum of a liquid solution phase. In a ternary or higher-dimensional system, using fine-meshed grids to map the liquid phase can become costly. We will therefore start with a low-resolution grid of $N_0=30$ gridpoints in each direction of the reduced mole fraction vector $\mathbf{\bar x}=(x_{\mathrm{CaO}},x_{\mathrm{Al}_2\mathrm{O}_3})$, subject to the constraint $0\le x_{\mathrm{CaO}}+x_{\mathrm{Al}_2\mathrm{O}_3}\le 1$. The third coordinate follows from $x_{\mathrm{SiO}_2}=1-x_{\mathrm{CaO}}-x_{\mathrm{Al}_2\mathrm{O}_3}$. We now compute the Gibbs free energies of all points in this 2-dimensional space, and thus produce a point cloud in 3D $(x_{\mathrm{CaO}},x_{\mathrm{Al}_2\mathrm{O}_3},G)$-space. For each of these points we keep additional information in storage, such as what the physical meaning of this point is: is it a \stoiphase{} point, and if so, which; is it a point of a continuum and if so, which: liquid or solid solution, etc. Then we feed this point cloud into the convex hull algorithm, which returns the indices of those points that are on the convex hull, as well as the simplices (here: triangles) that connect these points. The result for a temperature $T=1400^\circ$C=1673 K and a pressure of $P=1$ bar is shown in Fig.~\ref{fig-model-CaO-Al2O3-SiO2-lowres}. 

\subsection{Classification of simplices}
\label{sec-classification-of-simplices}
As one can see from the previous examples, the convex hull algorithm's task is essentially to triangulate the ternary plane into coexistence simplices\footnote{Fun fact: If one would replace the Gibbs free energy $\hatG$ with a parabolic function of $\mathbf{\bar x}$, one would obtain a Delaunay triangulation.}. The physical meaning of each of these simplices (indicated with color in Fig.~\ref{fig-model-CaO-Al2O3-SiO2-lowres}) is not given by the convex hull algorithm per se. They have to be determined aposteriori, once the triangulation is complete. In the \phasehull{} code, each simplex of the triangulation is inspected, and classified as one of the following:
\begin{itemize}
\item {All-fixed-composition-phase simplex:} If all three corner points are fixed-composition solid phases, then this simplex is a coexistence simplex of solids. In Fig.~\ref{fig-model-CaO-Al2O3-SiO2-lowres} these simplices are colored red/ros\'e.
\item {Two-fixed-one-liquid simplex:} If one of the three corner points is part of the liquid grid, while the others are fixed-composition phases, we have a coexistence of two fixed-composition solids with liquid melt. In Fig.~\ref{fig-model-CaO-Al2O3-SiO2-lowres} these simplices are colored purple/lilac.
\item {One-fixed-two-liquid tie line:} If two of the three corner points are part of the liquid grid, while only one corner point is a fixed-composition phase, we must investigate more carefully. Usually such simplices are 2D approximations of what, physically, should be a 1D tie line connecting a fixed-composition solid to the melt phase in equilibrium. Since a finite set of 1D lines cannot cover 2D space, the convex hull algorithm approximates these tie lines with narrow triangles, where the two liquid points are close to each other, while the fixed-composition phase is much farther away. For increasing grid resolution of the liquid, these ``tie line simplices'' become narrower, i.e., the ratio of the liquid-liquid distance to the liquid-fixedcomposition distance becomes smaller, and there will emerge an increasingly large number of them. However, there can be exceptional cases where a one-fixed-two-liquid simplex does not try to approximate a tie line. If the liquid-liquid distance is much larger than the typical liquid grid spacing, this is likely the case. In \phasehull{} this criterion is used to distinguish these cases. In the current example, however, all one-fixed-two-liquid simplices are tie lines.  In Fig.~\ref{fig-model-CaO-Al2O3-SiO2-lowres} they are colored green.
\item {Three-liquid simplices:} If all three corner points are part of the liquid grid, the simplex can have three possible physical meanings. If they are needle-like (narrow), like the one-fixed-two-liquid tie lines, then they are most probably tie lines covering a region of liquid immiscibility. Finally, if this simplex is tiny (i.e., it has the size of the local grid resolution), it is simply part of the discretized liquid phase. A sure-fire way to distinguish this from the previous two possibilities is that the Gibbs free energy function of the liquid at the barycenter of the simplex lies below the simplex. If it lies above, the simplex must be one indicating immiscibility. 
\end{itemize}

For binary systems this classification is simpler, while for quaternary or higher-dimensional systems, the classification can become somewhat complex. {When introducing solid solutions, in addition to the liquid and fixed-composition phases, the variety of possible simplices can become very large. It is, however, not always necessary to uniquely identify each simplex type separately. Often it suffices, for the purpose at hand, to only distinguish certain classes of simplex types, or to highlight only a few important ones. Even without any classification (e.g., Fig.~\ref{fig-model-CaO-Al2O3-SiO2-lowres} without the coloring), the geometry of the simplex-tiling of the phase diagram already contains useful information.}

\subsection{Adaptive grid refinement}
\label{sec-adaptive-grid-refinement}

\begin{figure*}
  \centerline{\includegraphics[width=\columnwidth]{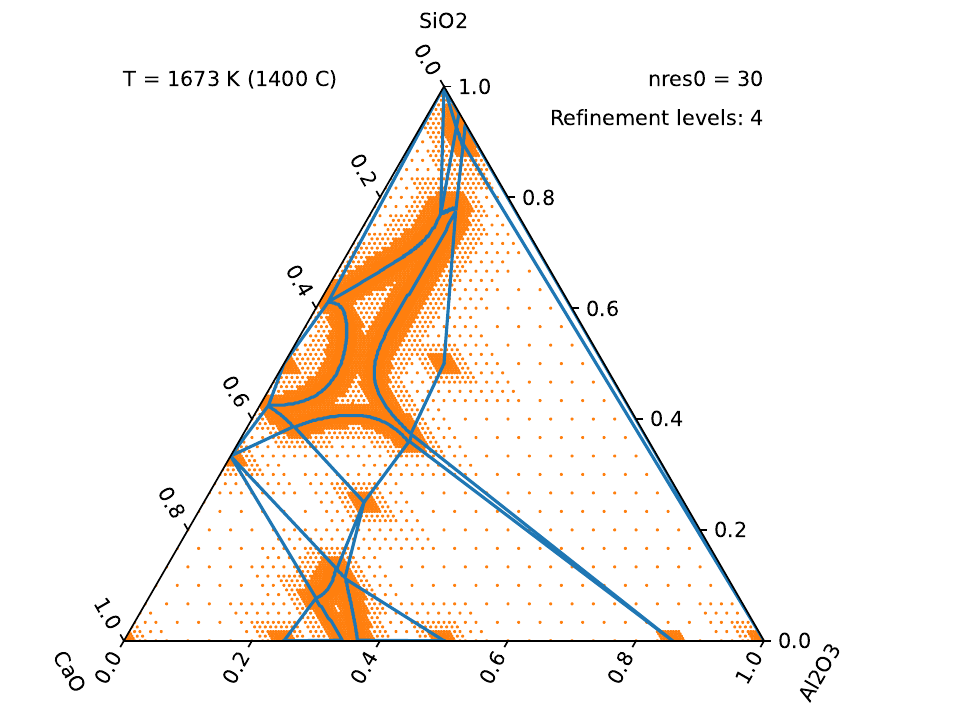}
  \includegraphics[width=\columnwidth]{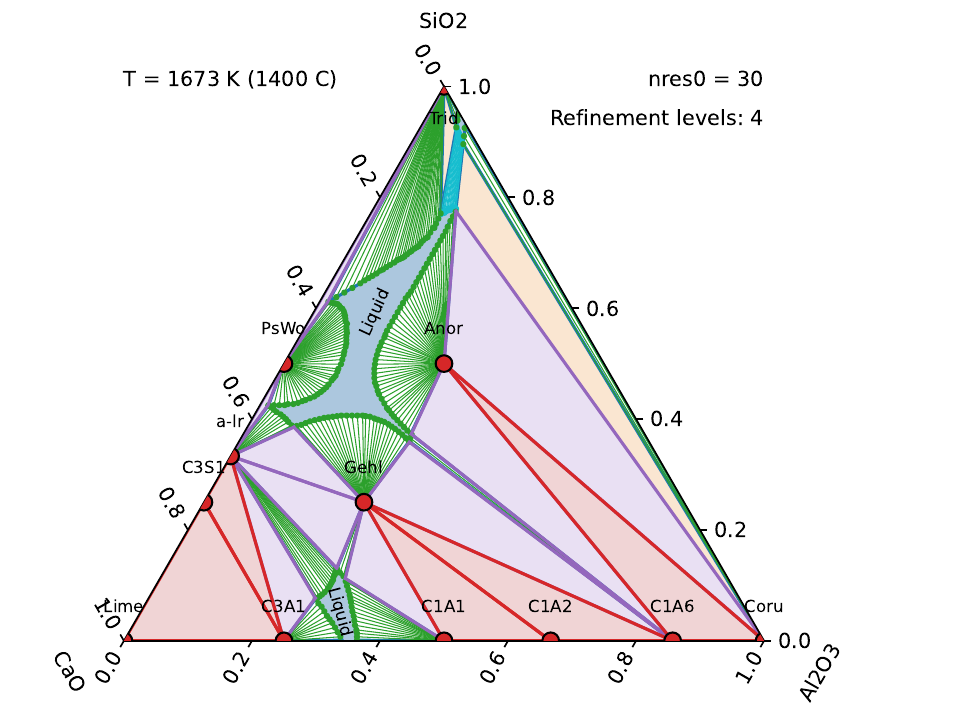}}
  \caption{\label{fig-model-CaO-Al2O3-SiO2-refined}Same model as Fig.~\ref{fig-model-CaO-Al2O3-SiO2-lowres}, but now with adaptive grid refinement. Left: The liquid grid points (orange points) after 4 levels of refinement from the base grid of $N_0=30$. Right: The resulting phase diagram at high resolution -- compare to Fig.~\ref{fig-model-CaO-Al2O3-SiO2-lowres}. A small region of liquid immiscibility emerged, covered with tie lines colored cyan.}
\end{figure*}

While the low-resolution gridding might be sufficient for many or most applications in geophysics and astronomy, it does not reveal details, and it looks suboptimal. However, we do not need high resolution everywhere. Only near a liquidus or a binodal line (where a tie line ends in the liquid phase) is higher grid resolution needed. In \phasehull{} this adaptive grid refinement is done in a staged manner. First a low-resolution model is constructed (such as shown in Fig.~\ref{fig-model-CaO-Al2O3-SiO2-lowres}). Then additional grid points are added at double the current resolution around points that are on the liquidus/binodal. Using Python's built-in {set} functionality it is easy to add grid points without duplication, especially when using integer mapping. Integer mapping avoids round-off problems when using sets. One can do this by multiplying $\mathbf{\bar x}$ by the desired highest resolution $N=2^qN_0$, where $q>1$ is the refinement level, to obtain integer coordinates. After the adaptive refinement one divides by $N$ again to obtain $\mathbf{\bar x}$). With the new refined grid we re-compute the Gibbs free energies (which is only needed at the new grid points, but in \phasehull{} it is done everywhere for convenience, because this is not the computational bottleneck). Then this new set of points is fed again into the convex hull algorithm, and a new, refined, phase diagram is obtained. We can do any number of such cycles of refinement until the desired precision is reached. In Fig.~\ref{fig-model-CaO-Al2O3-SiO2-refined}-left, the refined grid is shown for four cycles of refinement and a base grid with $N_0=30$. The model thus has a resolution equivalent to a grid with $N=2^4\times 30=480$, but with far fewer gridpoints, and thus computationally far cheaper, than if no adaptive grid refinement strategy were used. 

The resulting high-resolution phase diagram is shown in Fig.~\ref{fig-model-CaO-Al2O3-SiO2-refined}-right. For the liquid itself (the light-blue region) the grid lines (the edges of the triangles) are not plotted, as they are not physically meaningful anyway. And instead of plotting the full triangles of the tie-simplices, as in Fig.~\ref{fig-model-CaO-Al2O3-SiO2-lowres}, the actual tie lines from the fixed-composition phases to the middle of the two liquid points of the tie simplices are plotted in green, with green dots at each end. To avoid overcrowding, the green tie lines are plotted with a stride of 4, i.e., only every 4th tie line is shown.

Note that the tie lines occur in well-defined groups. In this case most groups belong clearly to one fixed-composition phase, meaning that the group describes the liquidus belonging to that phase. But in the small immiscible liquids region at the top, the tie lines connect binodal curves, and form a separate group. The \phasehull{} code sorts all tie line simplices into such groups. The tie lines of each group form contiguous bundles, where the tie simplices are neighbors along their long edges. 

\subsection{The CaAl$_2$Si$_2$O$_8$ -- NaAlSi$_3$O$_8$ -- KAlSi$_3$O$_8$ feldspar ternary system (Elkins \& Grove 1990)}

\begin{figure}
  \centerline{\includegraphics[width=1.1\columnwidth]{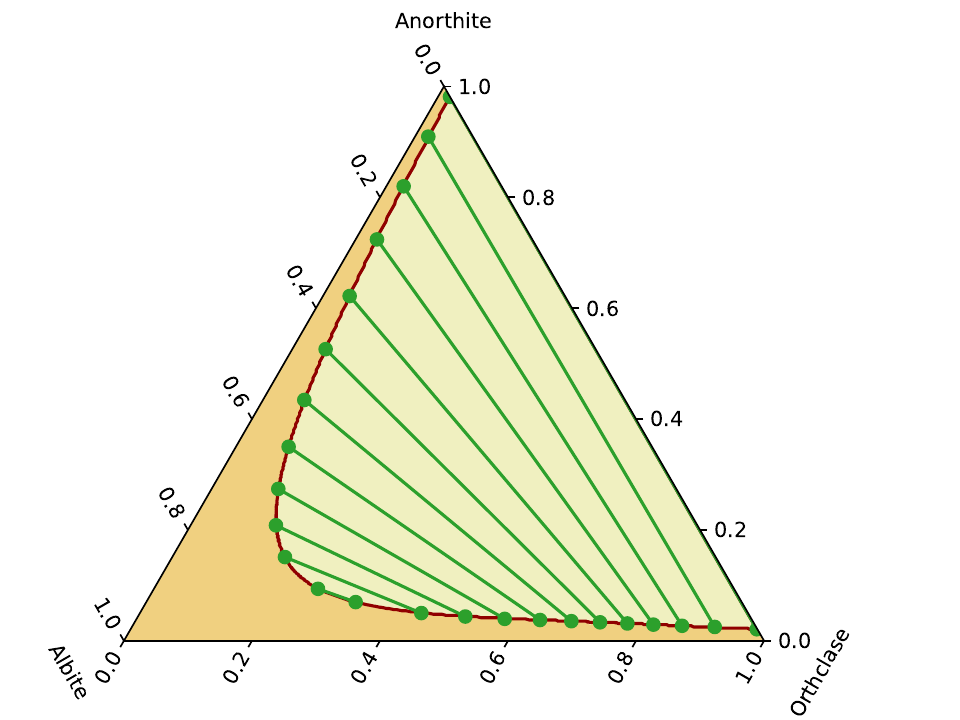}}
  \caption{\label{fig-model-feldspar}Phase diagram of the feldspar ternary system of \citet{ElkinsGrove:1990}, computed using the convex hull algorithm. Beige color: well-mixed phase. Green/yellow color: miscibility gap. Green: Tie lines.}
\end{figure}

The convex hull algorithm is particularly powerful for problems of solid or liquid solutions in ternary or higher-dimensional systems, where immiscibility can occur, and the tie lines lie on binodal curves. A well-known example is the Ca-Na-K feldspar system consisting of the components CaAl$_2$Si$_2$O$_8$, NaAlSi$_3$O$_8$, and KAlSi$_3$O$_8$. We follow the model of \citet{ElkinsGrove:1990}. It has no liquid phase or fixed-composition phases, just a continuum Gibbs function between the three components. It is convenient to set the Gibbs free energy at the corner points to zero, so that only the entropy terms and the excess Gibbs free energy via Margules parameters (their Table 4; see also appendix \ref{sec-precomputing-Gibbs}) come into play. 

The resulting phase diagram at 900$^\circ$C is shown in Fig.~\ref{fig-model-feldspar}. The computation of the binodal curve and the locations and directions of the tie lines follow reliably from the algorithm. 

\subsection{The high pressure MgSiO$_3$ -- Fe -- H$_2$ ternary system (Young et al.\ 2025)}

\begin{figure}
  \centerline{\includegraphics[width=1.1\columnwidth]{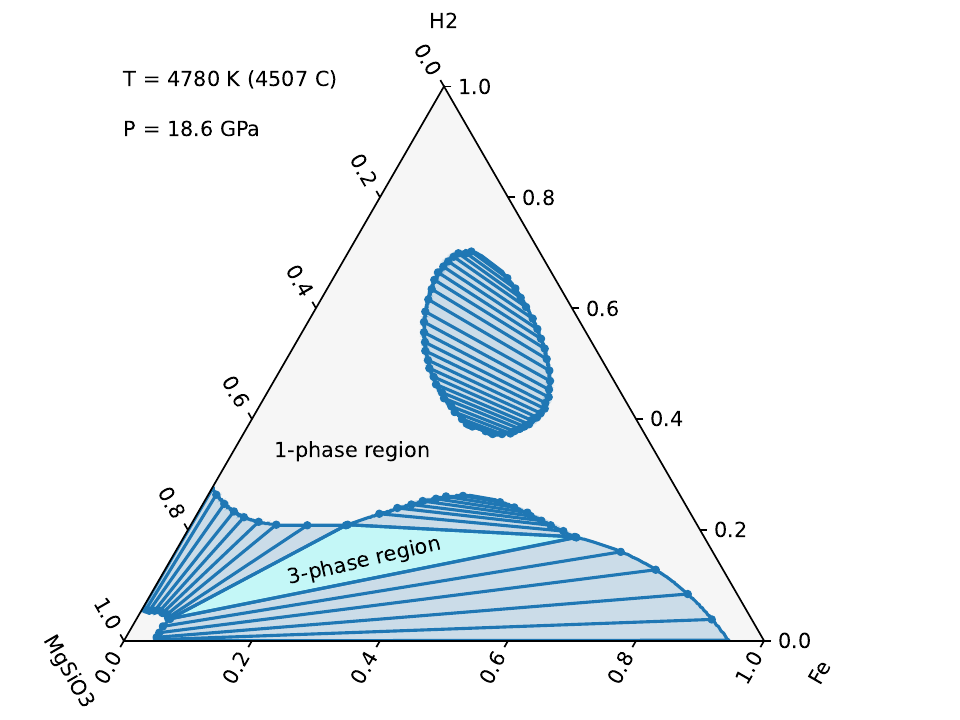}}
  \caption{\label{fig-model-subneptune}Phase diagram of the subneptune interior model of \citet{2025PSJ.....6..251Y}, at $T=4780$ K and $P=18.6$ GPa, computed using the convex hull algorithm. Background light grey color: well-mixed phase (1-phase region), light blue color: tie lines (2-phase region), cyan: tie triangle (3-phase region).}
\end{figure}

An example of a solution with rather complex behavior is the ternary system of liquid MgSiO$_3$, Fe and H$_2$ at high pressure and temperature, that was presented in \citet{2025PSJ.....6..251Y}. This model is meant to describe the (im)-miscibility of liquid phases in the deep interior of a sub-Neptune planet. The phase diagram shown in Fig.~\ref{fig-model-subneptune} shows two large regions of immiscibility. The bottom one consists, in itself, of four subregions, three of which are 2-phase regions covered by tie lines, and in the middle is a 3-phase region, which we call a ``tie triangle''. We refer to \citet{2025PSJ.....6..251Y} for the details of the model. Here it serves merely as a demonstration of the robustness of the \phasehull{} method.

\subsection{The Ag -- Cu binary alloy system (Chu et al.\ 2021)}

\begin{figure*}
  \centerline{\includegraphics[width=\columnwidth]{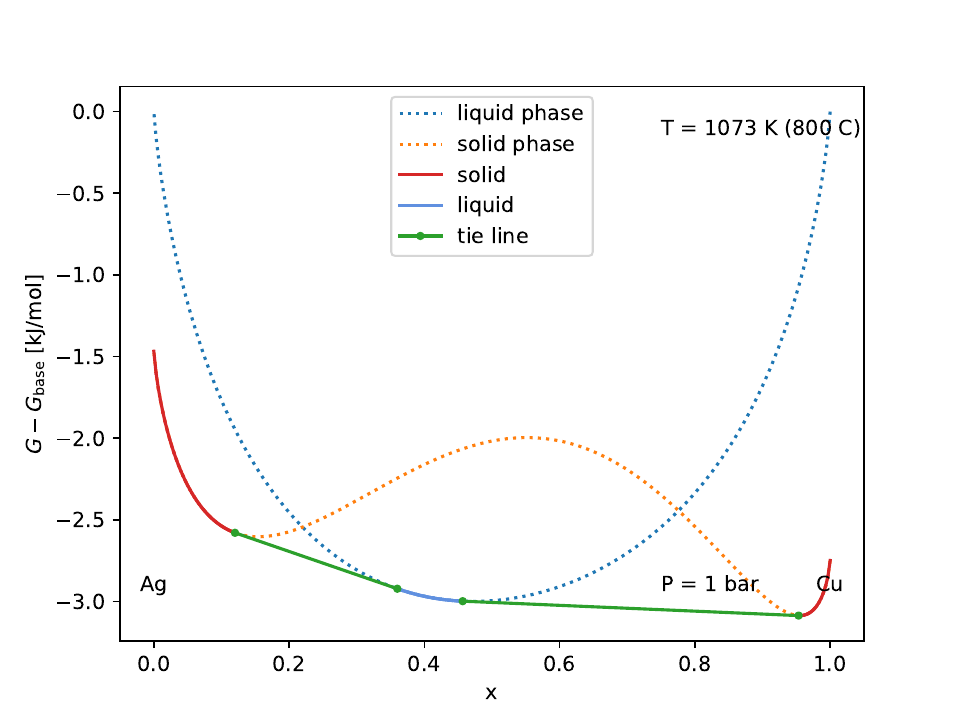}
  \includegraphics[width=\columnwidth]{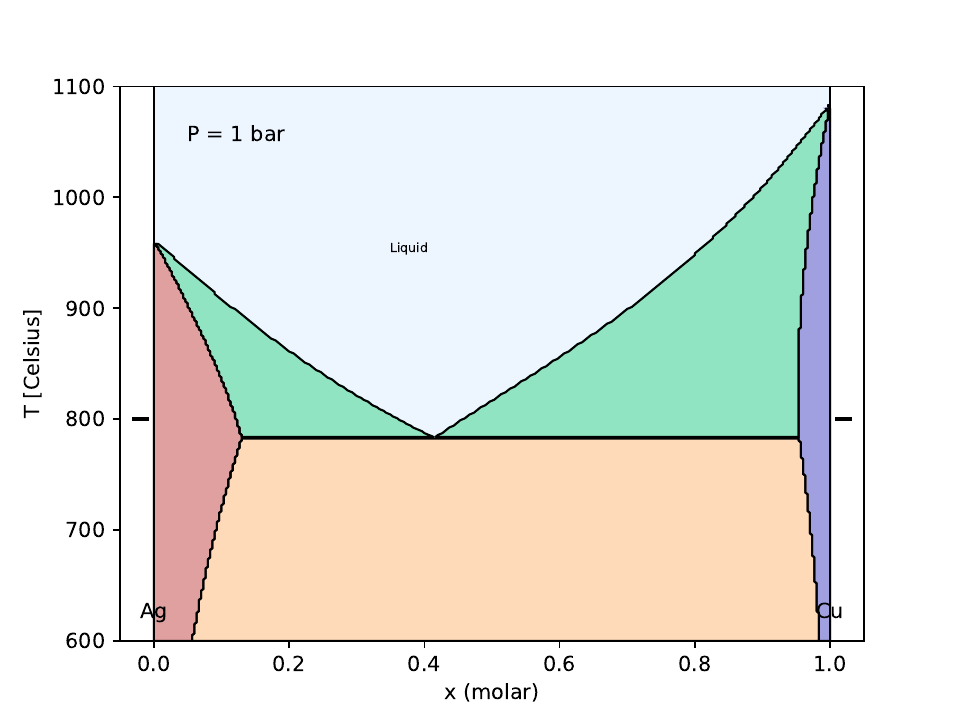}}
  \caption{\label{fig-model-AuCu}Model of an alloy of silver and copper. Left: Gibbs free energy diagram at $T=800^\circ$C, just above the eutectic temperature. In blue is the liquid phase, in red is the solid solution phase. In green the tie lines are shown that connect the liquid and the solid phases. Right: the $x$-$T$ phase diagram showing the full eutectic behavior of the system. The colors are as follows: Light blue is liquid, red-brown is solid silver mixed with some copper, blue is solid copper mixed with some silver, beige is the solid immiscibility region where both phases coexist, green are the two solid-liquid tie lines. The boundary between the beige and green regions is the eutectic isotherm. The boundary between the beige region and the red-brown/blue regions represent the solvus on each side. The boundary between the red-brown/blue regions and the green regions represent the solidus on each side. And finally, the boundaries between the green and light blue regions represent the liquidus. The markings on each side at 800 Celsius mark the temperature where the Gibbs free energy diagram (left panel) was computed.}
\end{figure*}

\phasehull{} can treat multiple continua simultaneously, such as a liquid solution coexisting with a solid solution. A well-studied system of this kind is the binary alloy of silver (Ag) and copper (Cu). Both the liquid and the solid phase exist as a continuously variable solution, with $x_{\mathrm{Ag}}+x_{\mathrm{Cu}}=1$. We therefore define a 1D model with two Gibbs functions, $\hatG^{\mathrm{sol}}(x)$ for the solid phase and $\hatG^{\mathrm{liq}}(x)$ for the liquid phase, with $x=x_{\mathrm{Cu}}=1-x_{\mathrm{Ag}}$. For the Gibbs functions we employ the model of \citet{CHU2021102233} in their limit of $r\rightarrow\infty$ (no nanoscale effects). Fig.~\ref{fig-model-AuCu}-left shows the Gibbs free energy at $T=800^\circ$C, demonstrating how the two continuous phases interact just above the eutectic temperature. Fig.~\ref{fig-model-AuCu}-right shows the $x-T$ phase diagram with the characteristic eutectic shape.

In \phasehull{} the primary difficulty of computing the $x-T$ phase diagram of Fig.~\ref{fig-model-AuCu}-right is not the computation itself, but the ``book keeping'' of the different regions across the different temperatures, in particular if the overall shape of the phase diagram is not known beforehand. For the $x-T$ diagram the MgO-SiO$_2$ binary system (Fig.~\ref{fig-model-MgO-SiO2-x-T}) this was easier, because each region was easily identifiable by the identity of the fixed-composition phase on either side of each region. Only if multiple immiscibility regions arise, we run into the same difficulty of identifying coherent regions across the different temperatures. In the case of alloys, none of the regions are easily identifiable by a label, so coherent regions have to be found by comparing the left and right $x$-values of regions between successive temperature values.

\subsection{Embedding a binary solid solution in a ternary phase diagram: The Diopside-Enstatite join.}
\label{sec-diopside-enstatite}

Within the ternary system of SiO$_2$, CaO, and MgO, there exist a binary solid solution between diopside (CaMgSi$_2$O$_6$) and enstatite (Mg$_2$Si$_2$O$_6$). One can write this as Ca$_{1-y}$Mg$_{y}$MgSi$_2$O$_6$, with $0\le y\le 1$ being the binary composition parameter. In other words, the diopside-enstatite join forms a continuous binary family of minerals. It covers a one-dimensional line within the two-dimensional SiO$_2$, CaO, and MgO ternary.

\begin{figure}
  \centerline{\includegraphics[width=1.1\columnwidth]{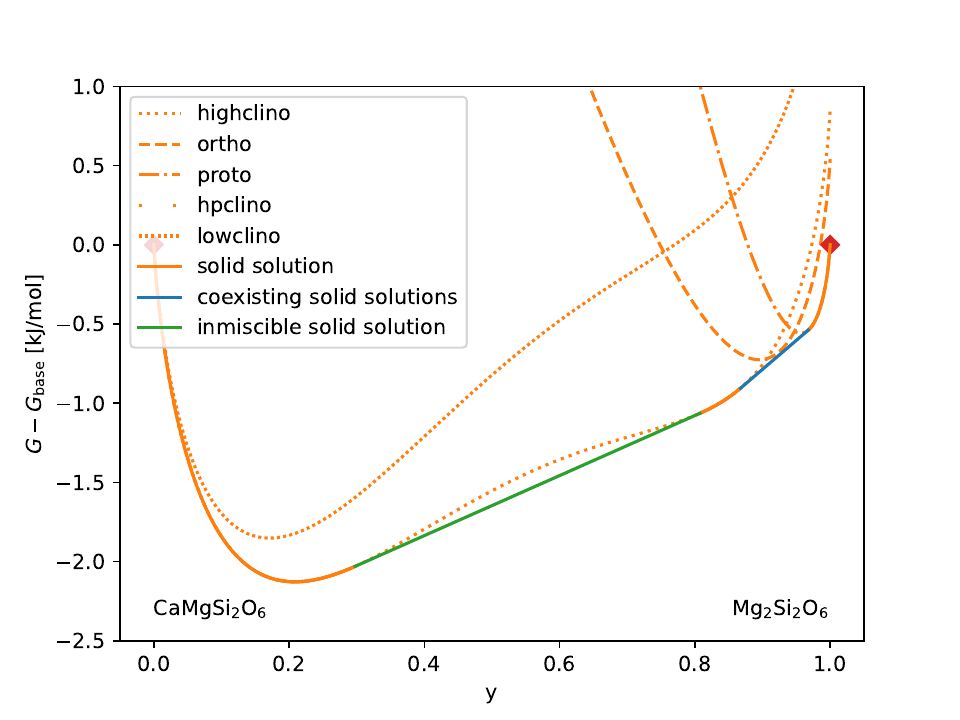}}
  \caption{\label{fig-dien-xG}The $y$-$\hat G$ diagram of the diopside-enstatite join according to \citet{gasparik:1990}. This model consists of five independent solid solutions: high-clino, low-clino, high-p-clino, proto and ortho. Also plotted are the stable phases as the bottom of the convex hull spanning all five solutions together.}
\end{figure}

In the Berman model of the SiO$_2$, CaO, and MgO system \citep{BermanPhD1983}, diopside and enstatite are only represented as fixed-composition minerals. So the Berman model does not account for the solid solution between diopside and enstatite. With \phasehull{} one can add the diopside-enstatite solid solution to the Berman model. For that we need a detailed thermodynamic model of this join. We use the model of \citet{gasparik:1990}, which includes five different solid solutions, each representing a different crystal structure. In Fig.~\ref{fig-dien-xG} we show the Gibbs free energies of these five binary solid solutions, and the corresponding tie lines, for temperature $T=1380$ C = $1653$ K and a pressure of 1 bar. One sees that the dominant solution for $y\lesssim 0.9$ is the ``high clino'' solution, which has a miscibility gap. But for $y\gtrsim 0.9$ the ``proto'' solution takes over. There is a tiny region (not visible in this figure) around $y\simeq 0.93$ where ``ortho'' briefly emerges, and so the blue line in Fig.~\ref{fig-dien-xG} is actually two tie lines, one between high clino and ortho, and one between ortho and proto.

\phasehull{} allows us to embed this one-dimensional model into the Berman ternary model, using the {\small\tt SolidSolution} class. We must take care to ensure mutual compatibility between the Berman and Gasparik models. Our choice, which is not guaranteed to be the best, is to calibrate the $\mu_0^{\mathrm{diop}}$ of highclino-diopside and the $\mu_0^{\mathrm{enst}}$ of proto-enstatite to the values of the fixed-composition diopside and enstatite from the Berman model, respectively. Furthermore, we must account for the fact that the scaled mineral formula units of diopside and enstatite, as used in the ternary phase diagram, are Ca$_{1/4}$Mg$_{1/4}$Si$_{1/2}$O$_{3/2}$ and Mg$_{1/2}$Si$_{1/2}$O$_{3/2}$, while the phase components that are mixed in the solid solution model of \citet{gasparik:1990} are CaMgSi$_2$O$_6$ and Mg$_2$Si$_2$O$_6$, which are four time multiples of the scaled phases. With the appropriate settings, \phasehull{} accounts for this automatically. 

\begin{figure*}
  \centerline{\includegraphics[width=0.67\columnwidth]{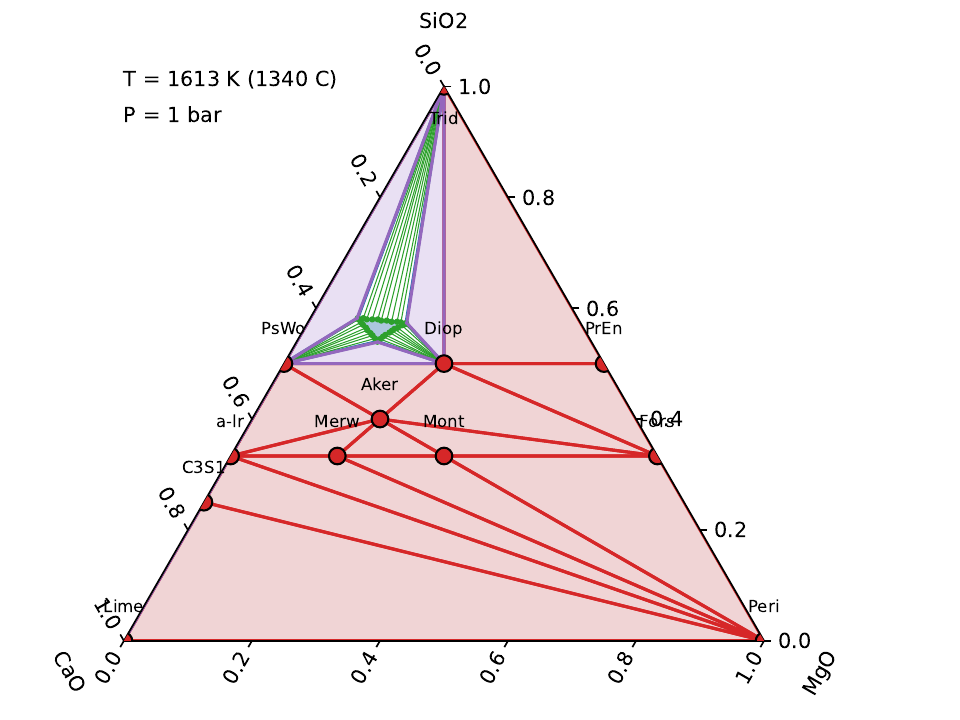}
              \includegraphics[width=0.67\columnwidth]{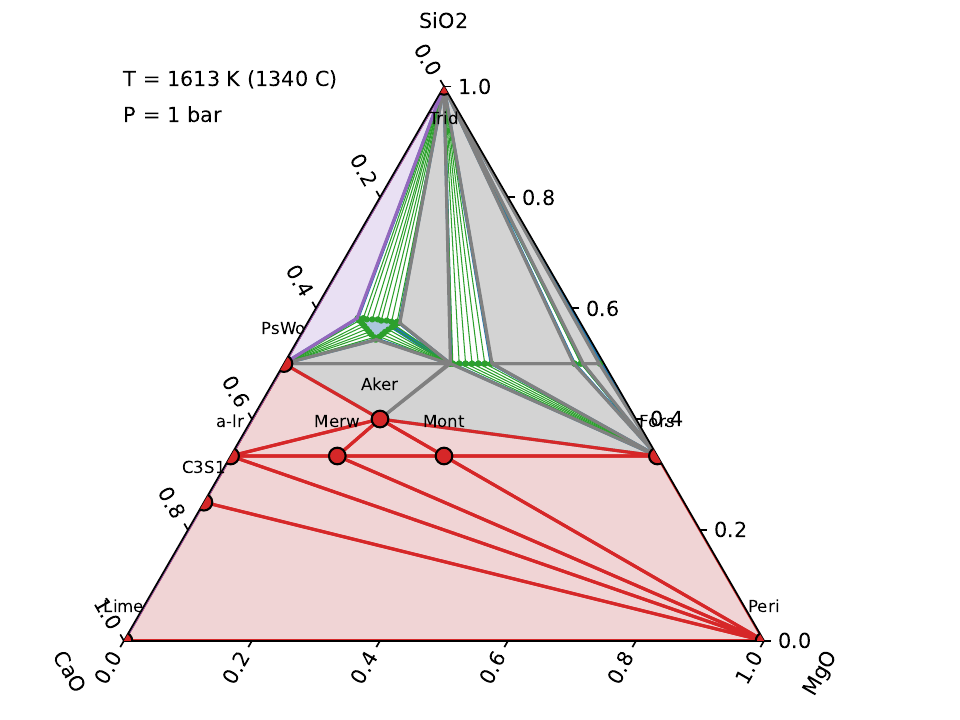}
              \includegraphics[width=0.6\columnwidth]{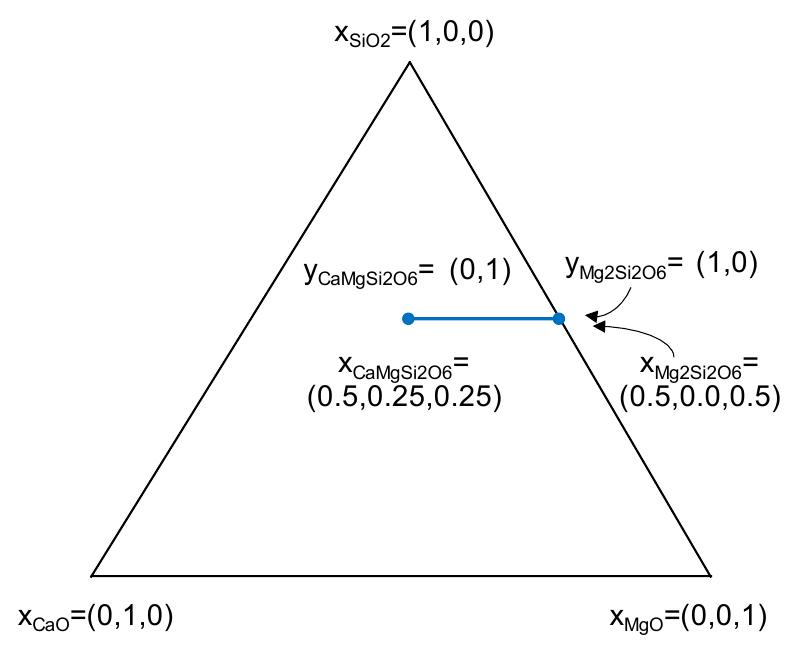}
  }
  \caption{\label{fig-dien-tern}The SiO$_2$, CaO, MgO ternary phase diagram at $T=1340$ C = $1613$ K and $P=1$ bar. Left: according to the model of \citep{BermanPhD1983} as implemented in \phasehull{}. Middle: The same ternary but now with the diopside-enstatite solid solution of \citet{gasparik:1990} embedded between the diopside and enstatite points. Right: Representation of the embedding of the solid solution (blue) in the ternary.}
\end{figure*}

The resulting ternary at $T=1340$ C = $1613$ K and $P=1$ bar is shown in Fig.~\ref{fig-dien-tern}, middle panel. On the left the ternary without the diopside-enstatite solid solution is shown to show the difference. 

The right panel shows how the diopside-enstatite solid solution is embedded in the ternary. We chose a fine grid for this solid solution (101 grid points between both ends), much finer than that of the liquid (base grid of 31 points, but with refinement near the liquidus). Both grids are independent. One of the results of this embedding is to modify the tie lines from the top (from silica) and the bottom (from forsterite). The miscibility gap along the join causes the tie lines to be split into two disjoint groups: those left of the miscibility gap and those right of it. However, the regions far away from the join are mostly unaffected, as is to be expected.

\subsection{Reciprocal solid solutions: The pyroxene quadrilateral phase diagram}
\label{sec-reciprocal-solutions}

{The simplest type of solid solution is a mixture of $N>1$ phase components, very much like a liquid (in the phase diagram sense). The diopside-enstatite join of Section \ref{sec-diopside-enstatite} is a binary example of that. However, solid solutions often behave in a more complex manner. For instance, if we expand the diopside-enstatite solid solution to include iron, we obtain the pyroxene quadrilateral phase diagram. Although this solid solution ``lives'' in a ternary space spanned by the phase components enstatite (Mg$_2$Si$_2$O$_6$), ferrosilite (Fe$_2$Si$_2$O$_6$), and wollastonite (Ca$_2$Si$_2$O$_6$), it covers only part of this ternary, namely the part for which $y_{\mathrm{Wo}}\le 0.5$. This is because the Mg, Fe and Ca atoms occupy the M2 (X) and M1 (Y) sites of the crystal, and only four (M2,M1) pairings (out of 9 in total) are energetically stable. These are: (Mg,Mg), (Fe,Fe), (Ca,Mg) and (Ca,Fe), i.e., enstatite, ferrosilite, diopside, and hedenbergite, respectively. Each unit of the crystal can thus only be one of these four pairings. The solid solution is therefore a solution of these four minerals, even though, from a compositional perspective, they are made up of only three independent components (enstatite, ferrosilite and wollastonite). This means that, for a given composition ${\bf y}=(y_{\mathrm{En}},y_{\mathrm{Fs}},y_{\mathrm{Wo}})$, there is one remaining internal degree of freedom that is not fixed by the composition. This freedom is characterized by the reciprocal reaction
\begin{equation}
  2\mathrm{Ca}\mathrm{Mg}\mathrm{Si}_2\mathrm{O}_6 + \mathrm{Fe}_2\mathrm{Si}_2\mathrm{O}_6
  \leftrightarrow
  2\mathrm{Ca}\mathrm{Fe}\mathrm{Si}_2\mathrm{O}_6 + \mathrm{Mg}_2\mathrm{Si}_2\mathrm{O}_6
\end{equation}
which says that two crystal units with (Ca,Mg) pairing and one crystal unit with (Fe,Fe) pairing can exchange ions and become two crystal units with (Ca,Fe) pairing and one crystal unit with (Mg,Mg) pairing, without changing the overall composition.}

\begin{figure*}
  \centerline{\includegraphics[width=0.60\columnwidth]{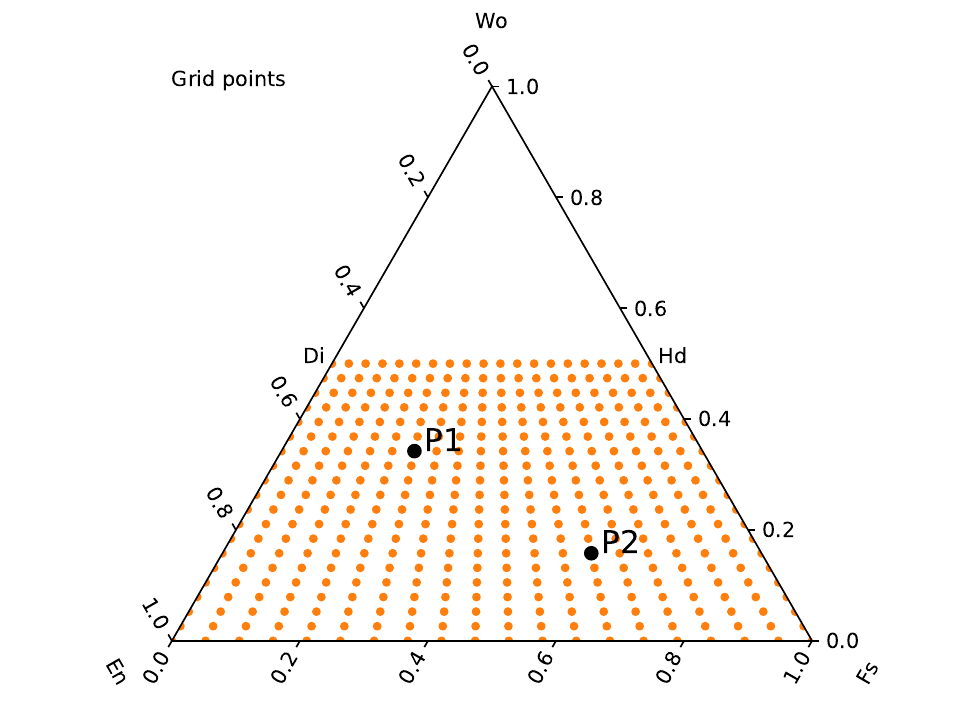}
              \includegraphics[width=0.67\columnwidth]{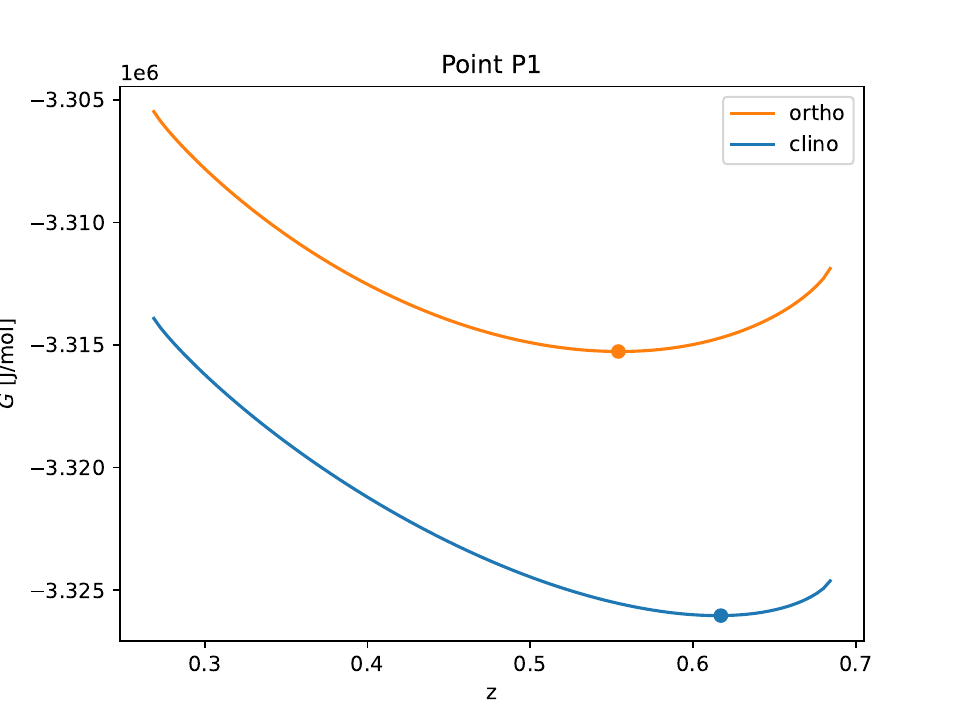}
              \includegraphics[width=0.67\columnwidth]{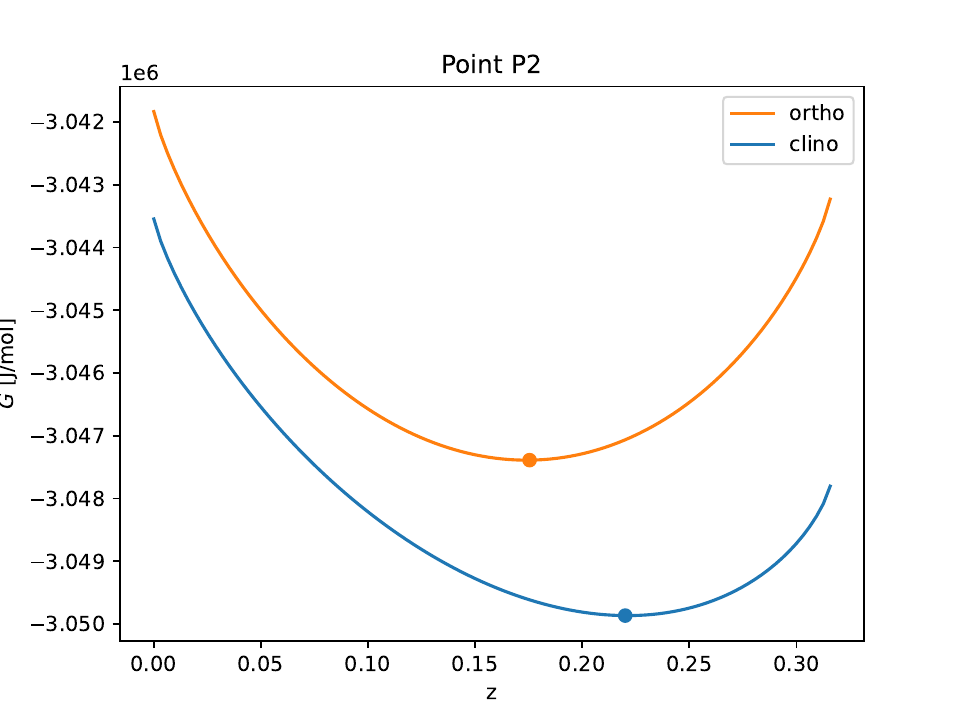}
  }
  \caption{\label{fig-pyrquadri-grid}{Left: The grid used for the pyroxene solid solution quadrilateral. Two points are selected for which, in the middle and right panel, the Gibbs free energy as a function of the internal degree of freedom $z$ (the reciprocal reaction) is shown for both solutions. The large dots in the middle and right panels are the points of lowest Gibbs free energy within the range of $z$. This is then the Gibbs free energy assigned to these points in the left panel. This is to demonstrate how, for each compositional point on the phase diagram, a unique $\hat G$ is defined, in spite of the internal degree of freedom $z$.}}
\end{figure*}

{Implementing such a reciprocal solid solution is straightforward in \phasehull{}. Like liquids, solid solutions are implemented as a grid. For the current example of the pyroxene solid solution living on the enstatite -- ferrosilite -- wollastonite ternary, we choose a grid that covers only that part of the ternary that lies within the quadrilateral spanned by enstatite (En), ferrosilite (Fs), diopside (Di) and hedenbergite (Hd), see Fig.~\ref{fig-pyrquadri-grid}. At each of these compositional grid points (except those at the very edge of the quadrilateral) the reciprocal reaction still leaves one internal degree of freedom. Let us call this $z$. One choice of the variable $z$ could be the abundance of diopside (or alternatively hedenbergite). For a given combination $(y_{\mathrm{En}},y_{\mathrm{Fs}},y_{\mathrm{Wo}},z)$ one can thus uniquely compute the corresponding mole fractions of the four mixed minerals $(\tilde y_{\mathrm{En}},\tilde y_{\mathrm{Fs}},\tilde y_{\mathrm{Di}},\tilde y_{\mathrm{Hd}})$. The question remains: For a given composition ${\bf y} = (y_{\mathrm{En}},y_{\mathrm{Fs}},y_{\mathrm{Wo}})$, which value should we take for $z$? The answer is: the value for which, at that composition, the Gibbs free energy is minimal (see Fig.~\ref{fig-pyrquadri-grid}-middle/right). This then uniquely determines the function $\hat G({\bf y})$.}

\begin{figure*}
  \centerline{\includegraphics[width=0.63\columnwidth]{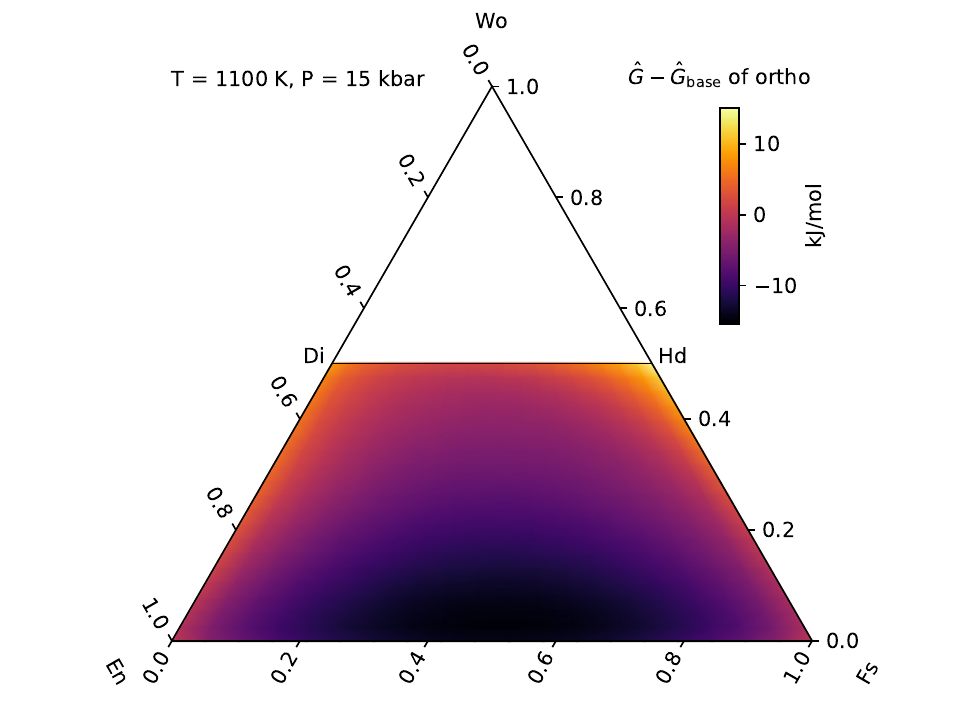}
              \includegraphics[width=0.63\columnwidth]{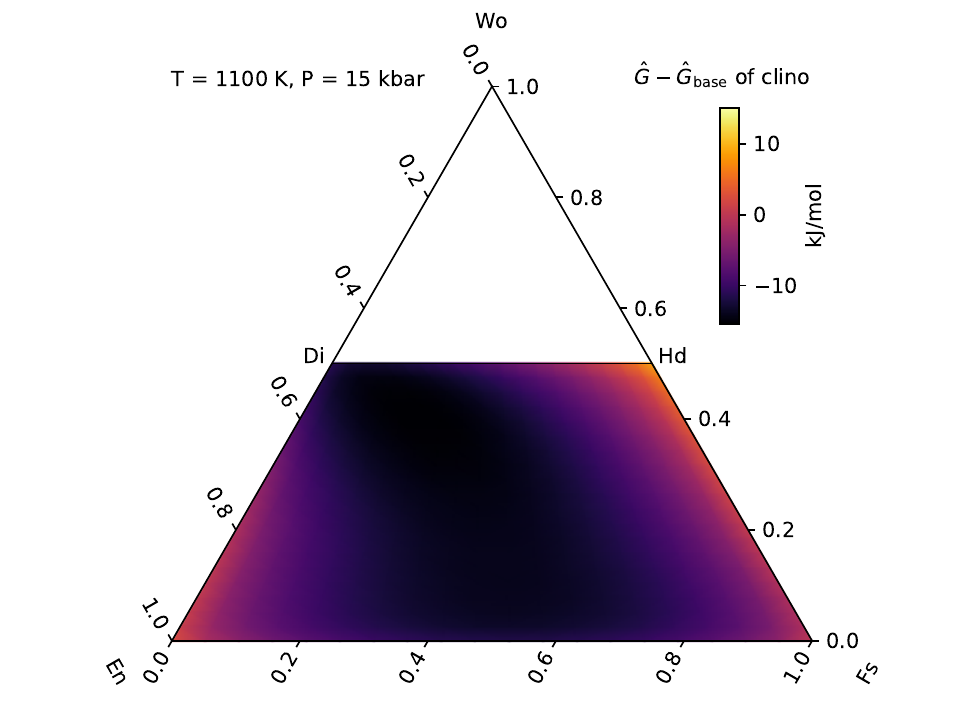}
              \includegraphics[width=0.63\columnwidth]{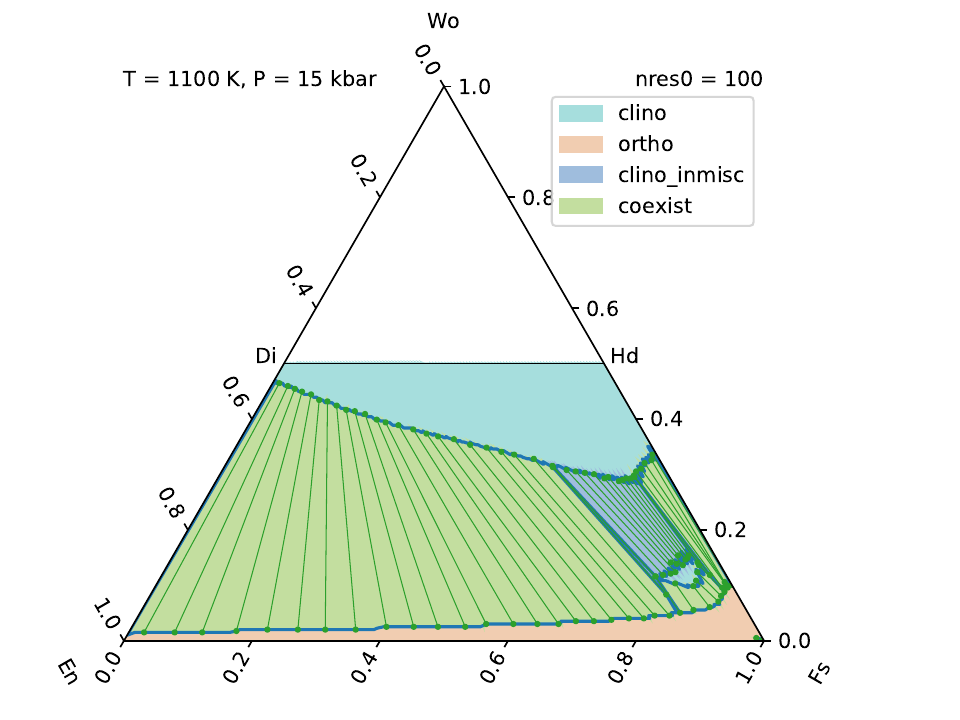}
  }
  \caption{\label{fig-pyrquadri-gibbs-and-diagram}{Left and middle panels: The Gibbs free energy $\hat G$ for the two solid solutions ortho and clino from the model of \citet{Saxena1986}. Right panel: The corresponding phase diagram computed with \phasehull{}.}}
\end{figure*}

{We demonstrate how this works for the model of \citet{Saxena1986}, which includes two solid solutions: ortho and clino. We implement both using the method above. The color maps of the Gibbs free energy $\hat G({\bf y})-\hat G_{\mathrm{base}}({\bf y})$ on the same (but arbitrary) color scale, are shown in Fig.~\ref{fig-pyrquadri-gibbs-and-diagram}-left/middle, for $T=1100$ K and $P=15$ kbar. The resulting phase diagram, computed with \phasehull{} (Fig.~\ref{fig-pyrquadri-gibbs-and-diagram}-right), can be compared to Fig.~6 of \citet{Saxena1982}. As expected, the top part of the quadrilateral consists of clinopyroxene, the bottom half of orthopyroxene, and both are connected with tie lines. To the right, the clinopyroxene develops a region of immiscibility. }

\subsection{Quaternaries and higher order systems}
\label{sec-quaternaries-higher}

\begin{figure*}
  \sidecaption
  \begin{minipage}[b]{12cm}
    \includegraphics[width=4.5cm]{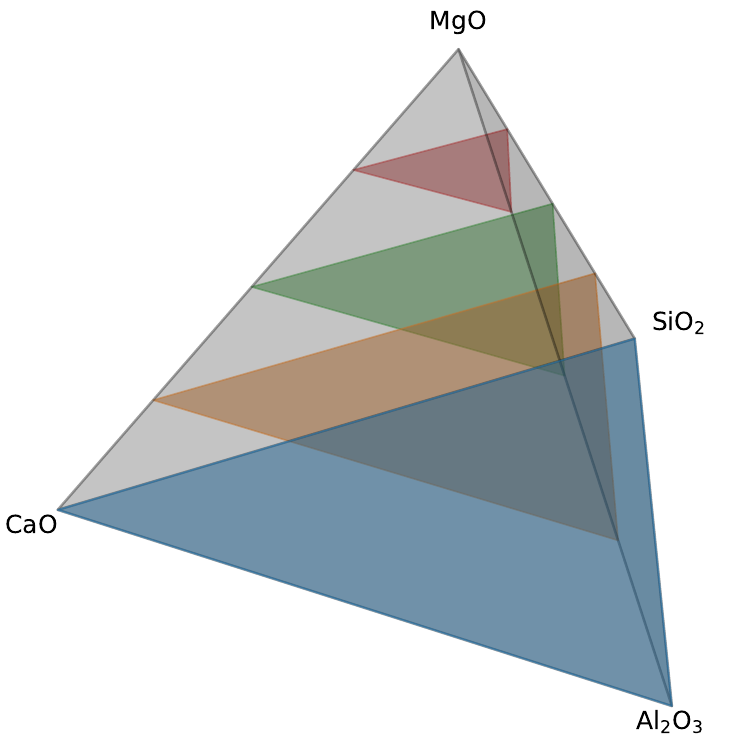}\hspace{1.3cm}\includegraphics[width=5.8cm]{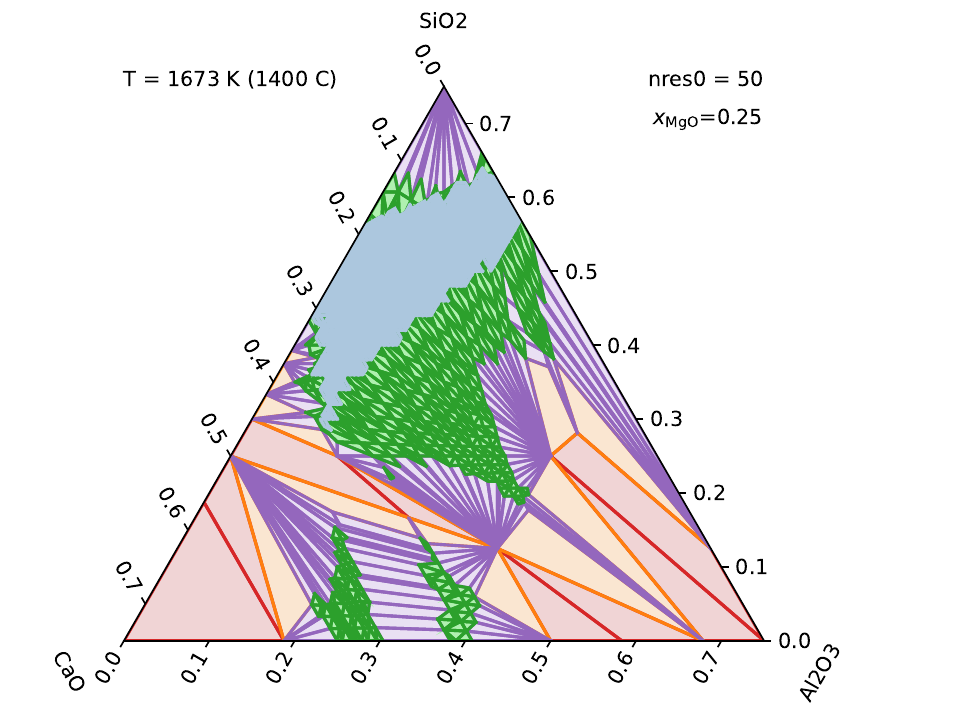}\\
    \includegraphics[width=5.8cm]{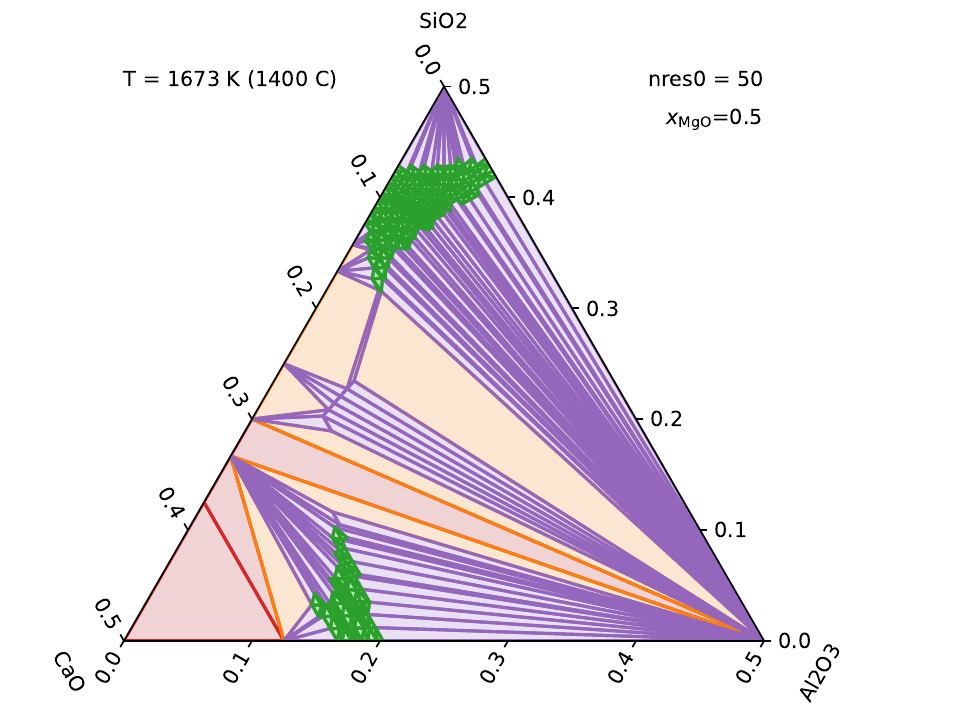}\includegraphics[width=5.8cm]{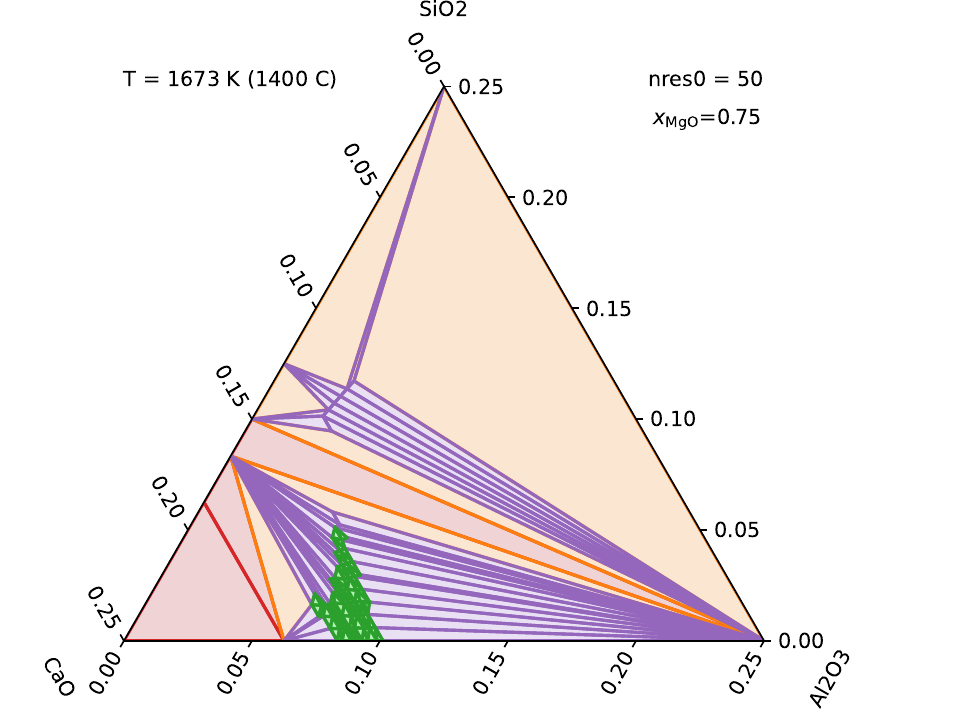}
  \end{minipage}
  \caption{\label{fig-quaternary-slices}{The SiO$_2$ -- CaO -- Al$_2$O$_3$ -- MgO quaternary phase diagram of the Berman 1983 model at $T=1400$ C = $1673$ K and $P=1$ bar. Top-Left: The slices taken. Top-right: the slice at $x_{\mathrm{MgO}}=0.25$, bottom-left: the slice at $x_{\mathrm{MgO}}=0.5$, bottom-right: the slice at $x_{\mathrm{MgO}}=0.75$. Note that the case for $x_{\mathrm{MgO}}=0$ is not shown here, as it is already shown in Fig.~\ref{fig-model-CaO-Al2O3-SiO2-lowres}.}}
\end{figure*}

{So far we only presented example applications of binary ($M=2$) and ternary ($M=3$) systems. But more realistic examples may involve more components. In principle, the extension of the convex hull algorithm to quaternary ($M=4$) and higher-dimensional systems is straightforward. If only a relatively limited number of fixed-composition phases are present, the algorithm will still remain fairly efficient. With liquids present, however, one may run into the ``curse of dimensionality'' for large $M$, as liquids require a complete discretized mapping of the composition space. The method will eventually become too slow for practical use for $M\ge 5$.

Here we, again, present the model of \citet[][see Sections \ref{sec-mgo-sio2} and \ref{sec-cao-al2o3-sio2}]{BermanPhD1983}, this time for the full CaO -- Al$_2$O$_3$ -- SiO$_2$ -- MgO quaternary system. For simplicity we do not include grid refinement, and we keep the sampling grid fixed at a moderately low resolution of 50 sampling points per dimension. The model takes about 20 seconds to run on an M1 MacBook Pro.

We present the results by taking 2D slices of the quaternary phase diagram at constant MgO abundance. We show these slices as ternaries as shown in Fig.~\ref{fig-quaternary-slices}. These slices consist of polygons that are always convex, but not always triangular. Each polygon is the cross section of the slice with one of the simplices of the quaternary. Most of these polygons are triangular, but some have four corner points. Computing the corner points of the polygonal cross sections of an $(M-1)$-dimensional simplex with a 2-dimensional hypersurface is a linear computational geometry problem that has to be solved for every simplex obtained from the convex hull algorithm. Fortunately, it is  cheap to compute.

As one can see in Fig.~\ref{fig-quaternary-slices}, what used to be the green tie lines at $x_{\mathrm{MgO}}=0$ (see Fig.~\ref{fig-model-CaO-Al2O3-SiO2-lowres}), now become smaller green triangles at $x_{\mathrm{MgO}}=0.25$ in the top-right panel. This is because they are still tie lines, but now in full 3D. At $x_{\mathrm{MgO}}=0.25$ these tie lines have also a vertical component of their direction. The horizontal slice thus cuts the green tie lines, and we see this like an obliquely cut broom. The finite size of these green triangles arises due to the finite (moderately low) spatial resolution of the sampling grid (50 samples across). The purple triangles now appear similar to what the green tie lines used to be at $x_{\mathrm{MgO}}=0$. This is because the purple simplices represent, in 3D, fans of 2-dimensional ``tie sheets'' connecting two fixed-composition solids to the liquid phase. The horizontal slice cuts through those sheets, producing lines, which is why the purple polygons appear like tie lines for $x_{\mathrm{MgO}}>0$. For $x_{\mathrm{MgO}}=0$ the ``tie sheet'' happens to lie perfectly in the $x_{\mathrm{MgO}}=0$ plane, which makes the purple polygons 2-dimensional in Fig.~\ref{fig-model-CaO-Al2O3-SiO2-lowres}. The beige colored polygons are the cross sections of the simplices connecting three fixed-composition solids to a liquid. They are 2-dimensional for $x_{\mathrm{MgO}}>0$ (Fig.~\ref{fig-quaternary-slices}) as well as for $x_{\mathrm{MgO}}=0$ (Fig.~\ref{fig-model-CaO-Al2O3-SiO2-lowres}). Finally the blue (liquid) and red (solid) simplices behave the same as in the ternary model.}

\begin{figure*}
  \sidecaption
  \includegraphics[width=6cm]{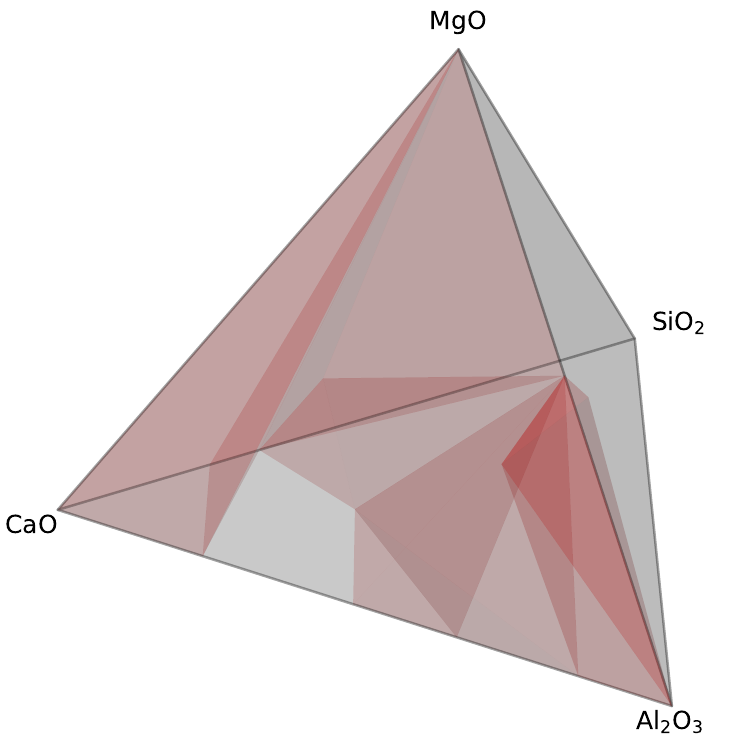}\includegraphics[width=6cm]{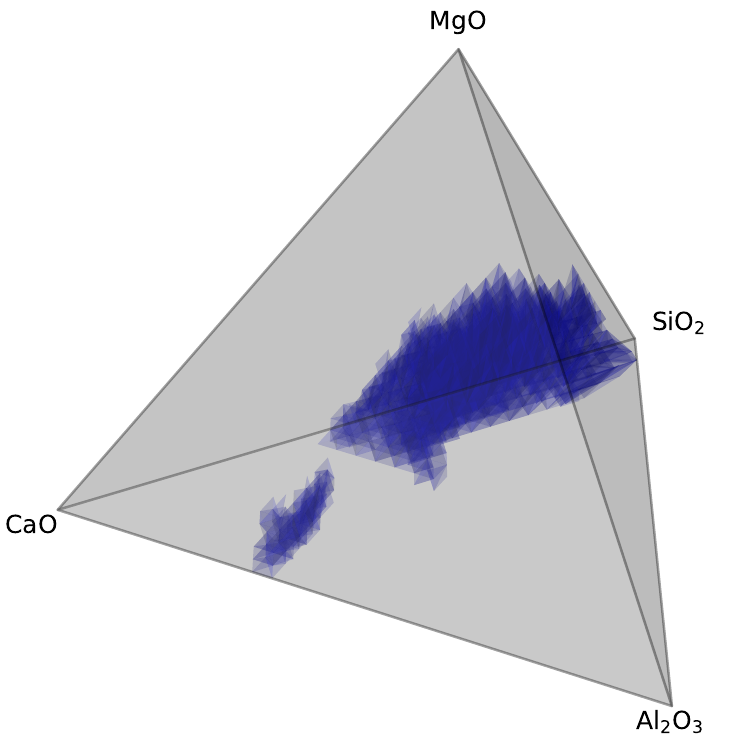}
  \caption{\label{fig-quaternary-volume}{The SiO$_2$ -- CaO -- Al$_2$O$_3$ -- MgO quaternary phase diagram of the Berman 1983 model at $T=1400$ C = $1673$ K and $P=1$ bar (same as Fig.~\ref{fig-quaternary-slices}). Left: all the simplices that are fully solid (having all four corners solid). Right: all the simplices that are fully liquid.}}
\end{figure*}

{Another way of visualizing the same quaternary phase diagram is directly in 3D as volume rendering, plotting only the simplices of a given type. This is shown in Fig.~\ref{fig-quaternary-volume} for all-liquid and all-solid simplices. Although it is not so easy to see the 3D geometry in static pdf format, in a live Python session one can rotate the 3D figure to get a better feeling for the geometry. This is particularly useful when viewing the more complex green (tie line) and purple (tie sheet) simplices, which are not shown here. If one is mainly interested in the coexisting phases for a single composition ${\bf x}$, a useful visualization could be to render the single simplex that the point ${\bf x}$ belongs to in 3D in a live Python session.}

\section{Discussion}
\label{sec-discussion}

\subsection{Strengths and weaknesses}
The algorithm that we present in this paper has its strengths, but also its weaknesses compared to conventional methods. In our opinion, its strengths are:
\begin{enumerate}
\item Although the classification of the simplices on the bottom of the convex hull (cf.\ Section \ref{sec-classification-of-simplices}) can produce errors, and the result may depend on the details of the resolution and grid-refinement in $x$-space, the convex hull itself is extremely reliable.
\item This reliability allows one to ``discover'' combinations of coexisting phases that one did not search for. The convex hull algorithm finds them automatically.
\item With sufficient grid refinement, the results are very accurate and stable, even at the usual double floating point precision.
\item The method is intuitive and simple, because it formulates the entire complexity of phase diagrams in terms of a single, well-understood mathematical problem: the search for the convex hull of a discrete set of points. And given that there are free software libraries to solve the convex hull problem accurately and reliably, the method is easy to implement.
\item For low temperature phase diagrams containing only fixed-composition phases, the method is exact. Each simplex represents an exact coexistence region in the phase diagram.
\end{enumerate}
These strengths make the method very useful for educational purposes, for the exploration of phase diagrams, and for accurate graphical representation of these phase diagrams. 

Its weaknesses include:
\begin{enumerate}
\item The accuracy of the location of the binodals depends entirely on the grid resolution in $x$-space. For applications that require high precision, the method quickly becomes computationally heavy.
\item The method only works when calculating, for a given temperature and pressure, the entire phase diagram. If one is interested only in the phase separation of a single bulk compositional point $x$, computing the entire phase diagram could become inefficient. This may be circumvented by pre-computing the phase diagrams, and using linear interpolation for individual queries, but that comes at the cost of either precision or computing memory.
\item For the computation of phases in systems with a large number of components, the method quickly becomes unfeasible, because each new component opens up a new dimension, so that the number of grid points scales exponentially with the number of components.
\item Computing phase diagrams in $(T,P)$-space can become computationally demanding, even for a moderate number of components, because a full phase diagram in ${\bf x}$-space has to be constructed for each set of $(T,P)$ values.
\item While the convex hull algorithm quickly and reliably computes the triangulation of the phase diagram, the proper identification and book-keeping of these simplices can become cumbersome. To alleviate this, the \phasehull{} code provides numerous methods to do this task in an automated way.
\end{enumerate}

\subsection{Similarity to the linear programming method of Connolly (2005)}

{One of the key features of the convex hull method presented in this paper is the precalculation of phases at a given temperature and pressure. For fixed-composition phases this is cheap, but for solutions (solid or liquid) this requires a grid of sampling points in ${\bf x}$-space. This can be expensive, but it has the advantage that the problem becomes completely linear.

This idea has been used before. It is, for instance, adopted by the widely-used Perple\_X software suite\footnote{\url{https://www.perplex.ethz.ch}} for computing phase equilibria and phase diagrams \citep{2005E&PSL.236..524C}. Perple\_X uses the linear programming algorithm called ``Simplex'' to minimize the energy of an assemblage at a single point ${\bf x}$ in composition space. For solution phases it adopts an identical approach to our method, sampling them over a grid of what it refers to as pseudo-compounds. Typically this method is used for fixed composition ${\bf x}$, and applied to a grid of values of temperature $T$ and pressure $P$.

In contrast, our method computes the phase diagram for all compositions ${\bf x}$ at once, but only for a given temperature $T$ and pressure $P$. And although the liquid and solid solution phases are represented by gridpoints, the resulting phase diagrams contain simplices that represent coexisting phases. For instance, even for super-high resolution gridding of the liquid phase, if the liquid coexists with a fixed-composition solid phase, the resulting simplex has a size that spans the full length of the tie line (or tie triangle) connecting the coexisting phases. The resulting phase diagram of our method thus returns the set of ``phase coexistence simplices'' of the phase diagram, not just a full sampling of the composition space. This becomes particularly apparent at low temperatures where no liquid phase exists: if all solids are represented by discrete fixed-compositions, our method returns only a few large simplices connecting the solid phases (see e.g.\ Fig.~\ref{fig-model-CaO-Al2O3-SiO2-lowtemp}), rather than a grid of sampled compositional points.

The Perple\_X method has the advantage that it is more efficient than our method for computing pseudosections, i.e., $T-P$ phase diagrams at fixed composition. Conversely, our method has the advantage at fixed $T$ and $P$, that it yields the complete geometry of tie lines, tie surfaces, and other coexistence simplices in one go. }

\subsection{The \phasehull{} code}

The algorithm described in this paper is implemented in the code called \phasehull{}, which is publicly available on GitHub\footnote{\url{https://github.com/dullemond/PhaseHull}}. The code will likely continue to develop, and updates will be posted on this repository. This code heavily relies on the NumPy library \citep{harris2020array}, the SciPy library \citep{2020SciPy-NMeth} and the QuickHull algorithm \citep{quickhull:1996} built in to SciPy. Visualization is done using the MatPlotLib library \citep{Hunter:2007} and the ternary figures use the mpltern\footnote{\url{http://mpltern.readthedocs.io}} library \citep{yuji_ikeda_2024_11068993}. \phasehull{} is designed to let these libraries do the heavy lifting, so that the code is reasonably fast in spite of being programmed in Python.

\section{Conclusion}
The convex hull algorithm, as implemented in the publicly available code \phasehull{}, is a powerful new method for astronomy and geophysics for computing {compositional} phase diagrams of solids, liquids and vapors {(i.e., phase diagrams in composition space)}. If provided with reliable Gibbs free energies for the material phases, the method computes the phase diagram by computing the convex hull of all potential phases in ${\mathbf x}-\hat G$ space. The bottom of this convex hull is then automatically the phase diagram, consisting of a set of simplices covering the full component space. All the complexity of the topology of the phase diagram, which would normally require a combination of various computations and tests, is handled fully and automatically by the external code/library that computes the convex hull. We chose the {\small\tt Qhull} library, as implemented in SciPy, which has been tested and improved since 1995. In this way \phasehull{} delegates all the complicated and demanding work to the reliable and efficient {\small\tt Qhull} package, minimizing the risk that programming errors affect the results. Only three tasks remain for \phasehull{} to do: (1) convert the physical problem into a set of $({\mathbf x},\hatG)$ points to be passed on to {\small\tt Qhull}, (2) ensure proper grid refinement where necessary, and (3) classifying the resulting simplices in terms of their physical meaning (i.e., if they are a tie line, or a 3-phase triangle, etc). The \phasehull{} package does include some Gibbs free energy databases to get started, but it is not meant as a model per se. It is meant as a tool: the user is supposed to provide his/her own Gibbs free energy databases and/or models, and can then use \phasehull{} to compute the corresponding phase diagrams.

\begin{acknowledgements}
We thank the organizers and sponsors of the Gordon Research Conference ``Origins of Solar Systems -- Constraints on Planet Formation from Theory, Observations, and Meteoritics'', chaired by M.~Sch\"onb\"achler and S.~Andrews, June 15-20, 2025, where part of this work was initiated. E.D.Y.\ acknowledges financial support from NASA grant 80NSSC24K0544 (Emerging Worlds program). C.P.D.\ thanks Drishika Nadella for her help with setting up the mineral databases, and Yves Marrocchi for inspiring discussions in April/May 2025 that triggered the present work. Finally, we like to thank the anonymous referee for his/her competent and helpful report that helped to improve the paper and put it into better context.
\end{acknowledgements}

\bibliographystyle{apj}
\bibliography{aa59967-26}

@Article{aflowchull_oses_2018,
author = {Oses, Corey and Gossett, Eric and Hicks, David and Rose, Frisco and Mehl, Michael J. and Perim, Eric and Takeuchi, Ichiro and Sanvito, Stefano and Scheffler, Matthias and Lederer, Yoav and Levy, Ohad and Toher, Cormac and Curtarolo, Stefano},
title = {AFLOW-CHULL: Cloud-Oriented Platform for Autonomous Phase Stability Analysis},
journal = {Journal of Chemical Information and Modeling},
volume = {58},
number = {12},
pages = {2477-2490},
year = {2018},
doi = {10.1021/acs.jcim.8b00393},
note ={PMID: 30188699},
URL = {https://doi.org/10.1021/acs.jcim.8b00393},
eprint = {https://doi.org/10.1021/acs.jcim.8b00393}
}

@ARTICLE{2010RJPCA..84..525V,
       author = {{Voskov}, A.~L. and {Voronin}, G.~F.},
        title = "{A universal method for calculating isobaric-isothermal sections of ternary system phase diagrams}",
      journal = {Russian Journal of Physical Chemistry A},
         year = 2010,
        month = apr,
       volume = {84},
       number = {4},
        pages = {525-533},
          doi = {10.1134/S0036024410040011},
       adsurl = {https://ui.adsabs.harvard.edu/abs/2010RJPCA..84..525V}
}

@ARTICLE{2000PhRvB..62.3648W,
       author = {{Widom}, Mike and {Al-Lehyani}, Ibrahim and {Moriarty}, John A.},
        title = "{First-principles interatomic potentials for transition-metal aluminides. III. Extension to ternary phase diagrams}",
      journal = {\prb},
         year = 2000,
        month = aug,
       volume = {62},
       number = {6},
        pages = {3648-3657},
          doi = {10.1103/PhysRevB.62.3648},
       adsurl = {https://ui.adsabs.harvard.edu/abs/2000PhRvB..62.3648W}
}

@ARTICLE{2008JMatR..23.1398G,
       author = {{Ghosh}, G.},
        title = "{Phase stability and cohesive properties of Au Sn intermetallics: A first-principles study}",
      journal = {Journal of Materials Research},
         year = 2008,
        month = may,
       volume = {23},
       number = {5},
        pages = {1398-1416},
          doi = {10.1557/JMR.2008.0175},
       adsurl = {https://ui.adsabs.harvard.edu/abs/2008JMatR..23.1398G}
}

@software{2013ascl.soft04016B,
       author = {{Barber}, C. Bradford and {Dobkin}, David P. and {Huhdanpaa}, Hannu},
        title = "{Qhull: Quickhull algorithm for computing the convex hull}",
 howpublished = {Astrophysics Source Code Library, record ascl:1304.016},
         year = 2013,
        month = apr,
          eid = {ascl:1304.016},
       adsurl = {https://ui.adsabs.harvard.edu/abs/2013ascl.soft04016B}
}

@ARTICLE{1995CoMP..119..197G,
       author = {{Ghiorso}, Mark S. and {Sack}, Richard O.},
        title = "{Chemical mass transfer in magmatic processes IV. A revised and internally consistent thermodynamic model for the interpolation and extrapolation of liquid-solid equilibria in magmatic systems at elevated temperatures and pressures}",
      journal = {Contributions to Mineralogy and Petrology},
         year = 1995,
        month = mar,
       volume = {119},
        pages = {197-212},
          doi = {10.1007/BF00307281},
       adsurl = {https://ui.adsabs.harvard.edu/abs/1995CoMP..119..197G}
}

@ARTICLE{1980CoMP...71..323G,
       author = {{Ghiorso}, M.~S. and {Carmichael}, I.~S.~E.},
        title = "{A regular solution model for met-aluminous silicate liquids: Applications to geothermometry, immiscibility, and the source regions of basic magmas}",
      journal = {Contributions to Mineralogy and Petrology},
         year = 1980,
        month = mar,
       volume = {71},
       number = {4},
        pages = {323-342},
          doi = {10.1007/BF00374706},
       adsurl = {https://ui.adsabs.harvard.edu/abs/1980CoMP...71..323G}
}

@ARTICLE{2005E&PSL.236..524C,
       author = {{Connolly}, J.~A.~D.},
        title = "{Computation of phase equilibria by linear programming: A tool for geodynamic modeling and its application to subduction zone decarbonation}",
      journal = {Earth and Planetary Science Letters},
         year = 2005,
        month = jul,
       volume = {236},
       number = {1-2},
        pages = {524-541},
          doi = {10.1016/j.epsl.2005.04.033},
       adsurl = {https://ui.adsabs.harvard.edu/abs/2005E&PSL.236..524C}
}

@ARTICLE{1984GeCoA..48..661B,
       author = {{Berman}, Robert G. and {Brown}, Thomas H.},
        title = "{A thermodynamic model for multicomponent melts, with application to the system CaO-Al $_{2}$O $_{3}$-SiO $_{2}$}",
      journal = {\gca},
         year = 1984,
        month = apr,
       volume = {48},
       number = {4},
        pages = {661-678},
          doi = {10.1016/0016-7037(84)90094-2},
       adsurl = {https://ui.adsabs.harvard.edu/abs/1984GeCoA..48..661B}
}

@PhdThesis{BermanPhD1983,
       author = {{Berman}, Robert G.},
        title = "A thermodynamic model for multicomponent melts, with application to the system CAO-MGO-AL2O3-SIO2",
       school = {University of British Columbia},
         year = 1983,
          doi = {10.14288/1.0052362},
}

@ARTICLE{2003CoMP..145..492H,
       author = {{Holland}, Tim and {Powell}, Roger},
        title = "{Activity-composition relations for phases in petrological calculations: an asymmetric multicomponent formulation}",
      journal = {Contributions to Mineralogy and Petrology},
         year = 2003,
        month = jul,
       volume = {145},
       number = {4},
        pages = {492-501},
          doi = {10.1007/s00410-003-0464-z},
       adsurl = {https://ui.adsabs.harvard.edu/abs/2003CoMP..145..492H}
}

@article{CHU2021102233,
title = {Thermodynamic reassessment of the Ag–Cu phase diagram at nano-scale},
journal = {Calphad},
volume = {72},
pages = {102233},
year = {2021},
issn = {0364-5916},
doi = {https://doi.org/10.1016/j.calphad.2020.102233},
url = {https://www.sciencedirect.com/science/article/pii/S036459162030496X},
author = {M.Z. Chu and Y.Z. Qin and T. Xiao and W. Shen and T. Su and C.H. Hu and Chengying Tang}
}

@Book{brentfultz:2000,
  author =       {Fultz, Brent},
  title =        {Phase Transitions in Materials},
  publisher =    {Cambridge University Press},
  year =         {2000},
}

@InCollection{hautier:2014,
  author =       {Hautier, Geoffroy},
  title =        {Data mining approaches to high-throughput crystal structure and compound prediction},
  booktitle =    {Prediction and Calculation of Crystal Structures},
  publisher =    {Springer Nature},
  year =         {2014},
  editor =       {Sule Atahan-Evrenk and Alan Aspuru-Guzik},
  volume =       {345},
  series =       {Topics in current chemistry},
  pages =        {139-180},
}

@article{ONG2013314,
title = {Python Materials Genomics (pymatgen): A robust, open-source python library for materials analysis},
author = {Shyue Ping Ong and William Davidson Richards and Anubhav Jain and Geoffroy Hautier and Michael Kocher and Shreyas Cholia and Dan Gunter and Vincent L. Chevrier and Kristin A. Persson and Gerbrand Ceder},
journal = {Computational Materials Science},
volume = {68},
pages = {314-319},
year = {2013},
issn = {0927-0256},
doi = {https://doi.org/10.1016/j.commatsci.2012.10.028},
url = {https://www.sciencedirect.com/science/article/pii/S0927025612006295},
}

@ARTICLE{1987E&PSL..82..207F,
       author = {{Fegley}, B. and {Cameron}, A.~G.~W.},
        title = "{A vaporization model for iron/silicate fractionation in the Mercury protoplanet}",
      journal = {Earth and Planetary Science Letters},
         year = 1987,
        month = apr,
       volume = {82},
       number = {3-4},
        pages = {207-222},
          doi = {10.1016/0012-821X(87)90196-8},
       adsurl = {https://ui.adsabs.harvard.edu/abs/1987E&PSL..82..207F}
}

@article{MaoHillert:2006,
author = {Mao, Huahai and Hillert, Mats and Selleby, Malin and Sundman, Bo},
title = {Thermodynamic Assessment of the CaO–Al2O3–SiO2 System},
journal = {Journal of the American Ceramic Society},
volume = {89},
number = {1},
pages = {298-308},
doi = {https://doi.org/10.1111/j.1551-2916.2005.00698.x},
url = {https://ceramics.onlinelibrary.wiley.com/doi/abs/10.1111/j.1551-2916.2005.00698.x},
eprint = {https://ceramics.onlinelibrary.wiley.com/doi/pdf/10.1111/j.1551-2916.2005.00698.x},
year = {2006}
}

@article{TanShi:2024,
author = {Tan, Jing and Shi, Chenying and Liu, Yuling and Deng, Tengfei and Wu, Qing and Du, Yong},
title = {Thermodynamic descriptions of the CaO–Al2O3 and CaO–Al2O3–SiO2 systems over the whole composition and temperature ranges},
journal = {Journal of the American Ceramic Society},
volume = {107},
number = {9},
pages = {6388-6409},
doi = {https://doi.org/10.1111/jace.19895},
url = {https://ceramics.onlinelibrary.wiley.com/doi/abs/10.1111/jace.19895},
eprint = {https://ceramics.onlinelibrary.wiley.com/doi/pdf/10.1111/jace.19895},
year = {2024}
}

@Article{ElkinsGrove:1990,
  author =       {Elkins, L. T. and Grove, T. L.},
  title =        {Ternary feldspar experiments and thermodynamic models},
  journal =      {American Mineralogist},
  year =         {1990},
  volume =       {75},
  number =       {5-6},
  pages =        {544-559},
}

@ARTICLE{2023A&A...676A..52T,
       author = {{Timmermann}, Anina and {Shan}, Yutong and {Reiners}, Ansgar and {Pack}, Andreas},
        title = "{Revisiting equilibrium condensation and rocky planet compositions. Introducing the ECCOPLANETS code}",
      journal = {\aap},
         year = 2023,
        month = aug,
       volume = {676},
          eid = {A52},
        pages = {A52},
          doi = {10.1051/0004-6361/202244850},
archivePrefix = {arXiv},
       eprint = {2307.00914},
 primaryClass = {astro-ph.EP},
       adsurl = {https://ui.adsabs.harvard.edu/abs/2023A&A...676A..52T}
}

@software{yuji_ikeda_2024_11068993,
  author       = {Yuji Ikeda},
  title        = {yuzie007/mpltern: 1.0.4},
  month        = apr,
  year         = 2024,
  publisher    = {Zenodo},
  version      = {1.0.4},
  doi          = {10.5281/zenodo.11068993},
  url          = {https://doi.org/10.5281/zenodo.11068993},
}

@ARTICLE{quickhull:1996,
       author = {Barber, C.B. and Dobkin, D.P. and Huhdanpaa, H.T.},
        title = "{The Quickhull algorithm for convex hulls}",
      journal = {ACM Transactions on Mathematical Software},
         year = 1996,
       volume = {22},
        pages = {469-483},
}

@ARTICLE{2020SciPy-NMeth,
  author  = {Virtanen, Pauli and Gommers, Ralf and Oliphant, Travis E. and
            Haberland, Matt and Reddy, Tyler and Cournapeau, David and
            Burovski, Evgeni and Peterson, Pearu and Weckesser, Warren and
            Bright, Jonathan and {van der Walt}, St{\'e}fan J. and
            Brett, Matthew and Wilson, Joshua and Millman, K. Jarrod and
            Mayorov, Nikolay and Nelson, Andrew R. J. and Jones, Eric and
            Kern, Robert and Larson, Eric and Carey, C J and
            Polat, {\.I}lhan and Feng, Yu and Moore, Eric W. and
            {VanderPlas}, Jake and Laxalde, Denis and Perktold, Josef and
            Cimrman, Robert and Henriksen, Ian and Quintero, E. A. and
            Harris, Charles R. and Archibald, Anne M. and
            Ribeiro, Ant{\^o}nio H. and Pedregosa, Fabian and
            {van Mulbregt}, Paul and {SciPy 1.0 Contributors}},
  title   = {{{SciPy} 1.0: Fundamental Algorithms for Scientific
            Computing in Python}},
  journal = {Nature Methods},
  year    = {2020},
  volume  = {17},
  pages   = {261--272},
  adsurl  = {https://rdcu.be/b08Wh},
  doi     = {10.1038/s41592-019-0686-2},
}

@Article{harris2020array,
 title         = {Array programming with {NumPy}},
 author        = {Charles R. Harris and K. Jarrod Millman and St{\'{e}}fan J.
                 van der Walt and Ralf Gommers and Pauli Virtanen and David
                 Cournapeau and Eric Wieser and Julian Taylor and Sebastian
                 Berg and Nathaniel J. Smith and Robert Kern and Matti Picus
                 and Stephan Hoyer and Marten H. van Kerkwijk and Matthew
                 Brett and Allan Haldane and Jaime Fern{\'{a}}ndez del
                 R{\'{i}}o and Mark Wiebe and Pearu Peterson and Pierre
                 G{\'{e}}rard-Marchant and Kevin Sheppard and Tyler Reddy and
                 Warren Weckesser and Hameer Abbasi and Christoph Gohlke and
                 Travis E. Oliphant},
 year          = {2020},
 month         = sep,
 journal       = {Nature},
 volume        = {585},
 number        = {7825},
 pages         = {357--362},
 doi           = {10.1038/s41586-020-2649-2},
 publisher     = {Springer Science and Business Media {LLC}},
 url           = {https://doi.org/10.1038/s41586-020-2649-2}
}

@Article{Hunter:2007,
  Author    = {Hunter, J. D.},
  Title     = {Matplotlib: A 2D graphics environment},
  Journal   = {Computing in Science \& Engineering},
  Volume    = {9},
  Number    = {3},
  Pages     = {90--95},
  publisher = {IEEE COMPUTER SOC},
  doi       = {10.1109/MCSE.2007.55},
  year      = 2007
}

@ARTICLE{2025PSJ.....6..251Y,
       author = {{Young}, Edward D. and {Werlen}, Aaron and {Marcum}, Sarah P. and {Stixrude}, Lars and {Dullemond}, Cornelis P.},
        title = "{Differentiation, the Exception, Not the Rule: Evidence for Full Miscibility in Sub-Neptune Interiors}",
      journal = {\psj},
         year = 2025,
        month = nov,
       volume = {6},
       number = {11},
          eid = {251},
        pages = {251},
          doi = {10.3847/PSJ/ae1012},
archivePrefix = {arXiv},
       eprint = {2507.00947},
 primaryClass = {astro-ph.EP},
       adsurl = {https://ui.adsabs.harvard.edu/abs/2025PSJ.....6..251Y}
}

@Article{gasparik:1990,
  author =       {Gasparik, Tibor},
  title =        {A thermodynamic model for the enstatite-diopside join},
  journal =      {American Mineralogist},
  year =         {1990},
  volume =       {75},
  pages =        {1080-1091},
}

@ARTICLE{2026ApJ...999..178W,
       author = {{Werlen}, Aaron and {Young}, Edward D. and {Schlichting}, Hilke E. and {Dorn}, Caroline and {Shahar}, Anat},
        title = "{The Effects of Non-ideal Mixing in Planetary Magma Oceans and Atmospheres}",
      journal = {\apj},
         year = 2026,
        month = mar,
       volume = {999},
       number = {2},
          eid = {178},
        pages = {178},
          doi = {10.3847/1538-4357/ae434d},
archivePrefix = {arXiv},
       eprint = {2602.05917},
 primaryClass = {astro-ph.EP},
       adsurl = {https://ui.adsabs.harvard.edu/abs/2026ApJ...999..178W}
}

@ARTICLE{Saxena1982,
author  = {Saxena, S. K.},
title   = {Fictive component model of pyroxenes and multicomponent phase equilibria},
journal = {Contributions to Mineralogy and Petrology},
year    = {1982},
pages   = {345},
volume  = {78},
doi     = {10.1007/BF00398930}
}

@ARTICLE{Saxena1986,
author  = {Saxena, S. K. and Sykes, J. and Eriksson, G.},
title   = {Phase Equilibria in the Pyroxene Quadrilateral},
journal = {Journal of Petrology},
year    = {1985},
pages   = {843–852},
volume  = {27},
doi     = {10.1093/petrology/27.4.843}
}

\clearpage

\begin{appendix}

\section{Computing the Gibbs free energies for a given $T$ and $P$}
\label{sec-precomputing-Gibbs}
\label{sec-databases}

For the convex hull method to work, all Gibbs free energies have to be precalculated for the temperature $T$ and pressure $P$ of interest. The Gibbs free energy of a substance for a given temperature $T$ and pressure $P$ can be computed from the enthalpy and entropy
\begin{equation}\label{eq-gibbs-for-T-and-P}
G(T,P) = H(T,P) - T\,S(T,P)
\end{equation}
with 
\begin{equation}\label{eq-H-for-T-and-P}
\begin{split}
  H(T,P) =& H_f^0 + \int_{298.15}^T C_p(T) dT \\
  & + \int_1^P \left[V(T,P)-T\left(\frac{\partial V}{\partial T}\right)_P\right]dP
\end{split}
\end{equation}
where $H_f^0$ is the enthalpy of formation at the standard temperature of $T=298.15$ K and pressure $P=1$ bar, and
\begin{equation}\label{eq-S-for-T-and-P}
\begin{split}
  S(T,P) =& S^0 + \int_{298.15}^T \frac{C_p(T)}{T} dT - \int_1^P \left(\frac{\partial V}{\partial T}\right)_PdP
\end{split}
\end{equation}
with $S^0$ is the entropy at the standard temperature and pressure. These values can be found in tables in the literature. The $C_p(T)$ is the specific heat capacity at constant pressure, which is typically temperature dependent. The $V(T,P)$ is the molar volume. For ideal gases $V=RT/P$, where $R=8.314$ J/(mol K). In the literature, polynomial fitting functions are given for $C_p(T)$ and $V(T,P)$, with coefficient listed in tables, so that the computation of the integrals can be carried out analytically for any fixed-composition mineral of interest. The resulting Gibbs free energy values are usually written as $\mu^0_k$ for mineral $k$.

Note that the molar values for $H$, $S$, $G$ and $C_p$ computed from the coefficients and formulae in the literature are usually for 1 mole of {unscaled} formula unit (cf.\ Section \ref{sec-scaled-formula-units}). In the notation of Section \ref{sec-scaled-formula-units}, we should have written them as $\tilde H$, $\tilde S$, $\tilde G$ and $\tilde C_p$. They can then be easily scaled to ``per mole of component'' by dividing them by their scaling factor as in Eq.~(\ref{eq-scaling-gibbs}).

For liquid or solid solutions we need, in addition to the $\mu^0$ values at the corner points, also the Gibbs free energy in between these points, i.e., at any composition ${\bf x}$ between the apices of the phase diagram (for liquids) or between the endmember phases (for a solid solution). This requires a model of the mixing of these system/phase components. Such a model can be formulated as the sum of an ideal mixing part plus an excess Gibbs free energy $\hatG^{\mathrm{ex}}({\bf x})$ (cf.~Eq.\ref{eq-gibbs-liquid-solution} for liquids and Eq.~\ref{eq-gibbs-solid-solution} or similar for a solid solution). As for the $C_p(T)$ and $V(T,P)$ functions, useful formulae for this excess Gibbs free energy can be found in various papers. But given that $\hatG_{\mathrm{ex}}({\bf x},T,P)$ is a high-dimensional function, it is more challenging to find a proper fitting formula for it. The simplest recipe is:
\begin{equation}
\hatG^{\mathrm{ex}}({\bf x}) = \frac{1}{2}\sum_{i,j}W^{G}_{ij} x_i x_j
\end{equation}
where $W^{G}_{ij}$, for $i\neq j$, are the binary interaction parameters for the Gibbs free energy, which can be written as
\begin{equation}
W^{G}_{ij} = W^{H}_{ij} - T W^{S}_{ij} + (P-1) W^{V}_{ij}
\end{equation}
with $T$ in Kelvin and $P$ in bar. $W^{H}_{ij}$, $W^{S}_{ij}$ and $W^{V}_{ij}$ are independent of $T$ and $P$, and are listed in tables in the applicable literature. These are the binary Margules parameters, which describe symmetric, regular non-ideal mixing behavior. The model of \citet{1995CoMP..119..197G}, of the MELTS code, is of this type. More complex mixing behavior can be introduced by the more general Margules formalism with higher-order versions such as $W_{ijk}$, $W_{ijkl}$ etc.,\ \citep{1984GeCoA..48..661B}. The model of Berman's PhD thesis \citep{BermanPhD1983} for the CaO-SiO$_2$-MgO-Al$_2$O$_3$ mineral system uses Margules parameters up to rank 4:
\begin{equation}
\hatG^{\mathrm{ex}}({\bf x}) = \sum_{i\le j\le k\le l}W^{G}_{ijkl} x_i x_j x_k x_l
\end{equation}
Alternatively, one can introduce asymmetry using the asymmetry parameter formalism of \citet{2003CoMP..145..492H} or the Redlich-Kister formulation. Another much-used recipe to define an interaction energy $\hatG^{\mathrm{ex}}$ is by specifying a formula for the activity coefficient $\gamma_i$ of each component as a function of the $x_0,\cdots,x_{M-1}$. The excess Gibbs free energy then follows from
\begin{equation}\label{eq-gibbs-from-activity-coeff}
\hatG^{\mathrm{ex}}({\bf x}) = RT \sum_i x_i\ln(\gamma_i)
\end{equation}
Care must then be taken that the formulae for the activity coefficient of the different components are mutually consistent, as they are not independent of each other.

The \phasehull{} software package contains several datasets of mineral systems and solution models from the literature. All of the models shown in this paper (Section \ref{sec-examples}) are included in the package, and further datasets and solution models are going to be added in due time. The package is designed to make it relatively straightforward to add one's own datasets and solution models. When combining datasets, it should, however, be kept in mind that mutually consistent definitions of the Gibbs free energy of formation are used. Furthermore, even if matching definitions are used, different datasets tend to be slightly differently calibrated. Key features of a phase diagram rely on Gibbs free energy {differences} between phases, meaning that small calibration differences between databases can lead to large errors in the location of liquidi, solidi, solvi etc in the resulting phase diagrams.

\section{Post-processing of the phase diagrams}
The convex hull algorithm provides us with a set of simplices covering the full compositional space at a given $T$ and $P$. The \phasehull{} code assigns physical meaning to each of these simplices, so that it becomes clear which phases are at the corner points, and what is the physical type of the simplex (a tie line, a tie triangle, part of a liquid, etc). A graphical representation can then be made by plotting each simplex with a color appropriate for the physical type.

To obtain the phase properties of the system at {some specific bulk composition} ${\bf x}$, however, one must perform some postprocessing calculations. The phase composition at a point ${\bf x}$ within a simplex requires application of the ``lever rule'' (Subsection \ref{sec-app-lever-rule}). And in case of a liquid or solid solution, the activities of the components at that point requires computation of the derivatives of the excess Gibbs free energy (Subsection \ref{sec-app-activity-coefficients}). 

\subsection{Lever rule for a system of $M$ components}
\label{sec-app-lever-rule}
Consider a system of $M$ components, with mole fractional coordinates ${\mathbf x}=(x_{0},\cdots,x_{M-1})$. Of these coordinates, only $M-1$ are independent. We arbitrarily choose the last coordinate $x_{M-1}$ as the redundant one, obeying $x_{M-1}=1-\sum_{i=0}^{M-2}x_i$. We define $\mathbf{\bar x}$ as
\begin{equation}
\mathbf{\bar x} = (x_{0},\cdots,x_{M-2})
\end{equation}
where the redundancy is now eliminated. The convex hull algorithm divides this $M-1$-dimensional space up into simplices, each consisting of $M$ corners. Any point $\mathbf{\bar x}$ lies inside of one of these simplices, or on its boundary. If we know in which simplex this point lies, we can decompose point $\mathbf{\bar x}$ into a linear sum of the corner points of the simplex
\begin{equation}\label{eq-bf-x-in-y}
\mathbf{\bar x} = \sum_{k=0}^{M-1} y_k \mathbf{\bar x}_k
\end{equation}
where $\mathbf{\bar x}_k$, with $k=0\cdots M-1$, are the coordinates of the $M$ corner points of the simplex. Also $y_{0},\cdots,y_{M-1}$ are not linearly independent, with $y_{M-1}=1-\sum_{k=0}^{M-2}y_k$, and like with ${\mathbf x}$ we define a $\mathbf{\bar y}=(y_{0},\cdots,y_{M-2})$. One can then write Eq.~(\ref{eq-bf-x-in-y}) as
\begin{equation}
(\mathbf{\bar x}-\mathbf{\bar x}_{M-1}) = \sum_{k=0}^{M-2} y_k (\mathbf{\bar x}_k-\mathbf{\bar x}_{M-1})
\end{equation}
In matrix form this equation is
\begin{equation}
{\cal M} \mathbf{\bar y} = (\mathbf{\bar x}-\mathbf{\bar x}_{M-1})
\end{equation}
where the matrix ${\cal M}$ is constructed from the juxtaposition of the vectors $(\mathbf{\bar x}_k-\mathbf{\bar x}_{M-1})$. The coefficients $y_{0},\cdots,y_{M-2}$ are now found by matrix inversion
\begin{equation}
\mathbf{\bar y} = {\cal M}^{\mathrm{inv}}(\mathbf{\bar x}-\mathbf{\bar x}_{M-1})
\end{equation}
The value of $y_{M-1}$ is, finally, obtained from $y_{M-1}=1-\sum_{k=0\cdots M-2}y_k$, completing the set of coefficients needed for Eq.~(\ref{eq-bf-x-in-y}). The molar Gibbs free energy is then
\begin{equation}
\hat G = \sum_{k=0}^{M-1} y_k \hat G_k
\end{equation}
This is the $M$-dimensional lever rule. Note that $y$ is normalized as follows:
\begin{equation}\label{eq-y-normalization}
\sum_{k=0}^{M-1} y_k = 1
\end{equation}

\subsection{Activity coefficients}
\label{sec-app-activity-coefficients}
For non-ideal liquid or solid solutions, the activity $a_i$ of each of the components $i$ of the mixed system at bulk composition $x_i$ is often required for further chemical or physical computations. The activity coefficient $\gamma_i$ relates $x_i$ to $a_i$ according to $a_i = \gamma_i x_i$, where $\gamma_i$ is, itself, a function of $x_i$. For ideal mixing, $\gamma_i=1$. Any deviation of $\gamma_i$ from unity is an indication of non-ideality, caused by the excess Gibbs free energy $\hatG_{\mathrm{ex}}(x_0,\cdots,x_{M-1})$ and its derivatives. For a complete set of system components with $i=0,\cdots,M-1$, the formula is \citep[][their Eq.\ 15]{1984GeCoA..48..661B}:
\begin{equation}\label{eq-gamma-act-from-Gex}
RT\ln(\gamma_i) = \hatG_{\mathrm{ex}} + \sum_{n\neq i} x_n\left(\frac{\partial \hatG_{\mathrm{ex}}}{\partial x_{i,n}}\right)
\end{equation}
where the derivative operator $\partial/\partial x_{i,n}$ is a derivative along $dx_i$ and $dx_n$ with $dx_n=-dx_i$, and keeping all other $x$-components constant. The reason for this definition is that in taking the derivative with respect to $x_i$, it would be unphysical to keep all other $x$-components constant, because that would break the rule that $\sum_ix_i=1$. So when taking an infinitesimal step in direction $i$, we take the same step backward in direction $n$, so that $\sum_ix_i$ remains 1. For the Margules formulation of the excess Gibbs free energy $\hatG_{\mathrm{ex}}$, \citet{1984GeCoA..48..661B} give the general expression of Eq.~(\ref{eq-gamma-act-from-Gex}) in analytic closed form (their Eq.\ 22), which can directly be used in a computer code. This is the way it is implemented in \phasehull{}. But \phasehull{} also includes tools to evaluate Eq.~(\ref{eq-gamma-act-from-Gex}) through numerical differencing. For solution models in which the excess Gibbs free energy is defined through formulae for the activities in the first place (see Section \ref{sec-precomputing-Gibbs}, Eq.~\ref{eq-gibbs-from-activity-coeff}), the computation of these activities is immediate, and the application of Eq.~(\ref{eq-gamma-act-from-Gex}) is not needed.

\section{Equivalence of common tangents and equal chemical potentials formulations of equilibrium}
\label{sec-app-liquidus}

A tie line connects two co-existing phases which are in chemodynamic equilibrium with each other. The \phasehull{} software finds these tie lines by applying the convex hull algorithm. Whenever a tie line connects to a continuum (i.e., a solution) on one or both ends, it forms a tangent to that continuum (see Fig.~\ref{fig-cartoon-crystliquid}). If both ends connect to a continuum, the tie line forms a {common tangent} (see Fig.~\ref{fig-cartoon-continuous-phases}). This is how \phasehull{} finds equilibria between fixed-composition phases and liquids, between different compositions of the same liquid, between solid solutions and liquids, and so forth.

The equilibrium condition can, however, also be formulated as the chemical potentials of both phases being equal to each other. The common tangent formulation and the chemical potential formulation of chemodynamic equilibrium are equivalent and interchangeable. This equivalence is, however, somewhat subtle. Although it is textbook knowledge, for convenience we recapitulate it here, in order to better put the \phasehull{} approach into context. 

We choose the example of a fixed-composition solid phase of composition ${\bf x}_s$ in equilibrium with a liquid. The case for a common tangent between two continua works along the same line. The {liquidus} belonging to the fixed-composition solid phase at a given temperature and pressure is defined as the set of locations ${\bf x}_l$ where the line connecting the liquid with that fixed-composition phase is tangent to the Gibbs free energy function of the liquid (the right-side end of the green tie line in Fig.~\ref{fig-cartoon-crystliquid}). We work this out for a simple 1D binary system consisting of components A and B. The liquid forms of components A,B have molar Gibbs free energies
\begin{equation}
\hatG_{A,B}^{\mathrm{liq}} = \mu_{A,B}^{0,\mathrm{liq}}
\end{equation}
where $\hatG_{A,B}^{\mathrm{liq}}$ denotes the Gibbs free energy of 1 mole of pure liquid A or B and $\mu_{A,B}^{0,\mathrm{liq}}$ are the standard state Gibbs free energies of these pure liquids. For a liquid mixture of both components, the Gibbs free energy is:
\begin{equation}\label{eq-gibbs-liquid-afo-xab-act}
\begin{split}
  \hatG^{\mathrm{liq}}(x_A,x_B) =& x_A\mu_{A}^{0,\mathrm{liq}} + x_B\mu_{B}^{0,\mathrm{liq}} \\
 & + RT \left[ x_A\ln(a_A) + x_B\ln(a_B) \right]
\end{split}
\end{equation}
where $x_{A,B}$ are the mole fractions of components A,B in the liquid, obeying $x_{A}+x_{B}=1$, and
$a_{A,B}$ are the activities of these components. The activities are $a_{A,B} = \gamma_{A,B} x_{A,B}$
where $\gamma_{A,B}$ are the activity coefficients which are, themselves, functions of $x_{A,B}$. With Eq.~(\ref{eq-gibbs-from-activity-coeff}), Eq.~\ref{eq-gibbs-liquid-afo-xab-act} then becomes
\begin{equation}\label{eq-gibbs-liquid-afo-xab-gex}
  \begin{split}
  \hatG^{\mathrm{liq}}(x_A,x_B) =&  x_A\mu_{A}^{0,\mathrm{liq}} + x_B\mu_{B}^{0,\mathrm{liq}} \\
  &+ RT \left[ x_A\ln(x_A) + x_B\ln(x_B)\right] \\
  &+ \hatG_{\mathrm{ex}}(x_A,x_B)
  \end{split}
\end{equation}
For convenience we will replace $x_A$ and $x_B$ with $x$, defined as $x_A = x$ and $x_B = 1-x$.
Eq.~\ref{eq-gibbs-liquid-afo-xab-gex} then becomes
\begin{equation}\label{eq-gibbs-liquid-afo-x}
  \begin{split}
    \hatG^{\mathrm{liq}}(x) =& x\mu_{A}^{0,\mathrm{liq}} + (1-x)\mu_{B}^{0,\mathrm{liq}}\\
    &+ RT \left[ x\ln(x) + (1-x)\ln(1-x) \right] \\
    &+ \hatG_{\mathrm{ex}}(x)
  \end{split}
\end{equation}

Now let us introduce a fixed-composition solid S with stoichiometry coefficients $\nu_A$ and $\nu_B$, meaning that a single formula unit can form from  $\nu_A$ units of component A and $\nu_B$ units of component B:
\begin{equation}
S \leftrightarrow \nu_A\, A + \nu_B\, B 
\end{equation}
For the example of forsterite Mg$_2$SiO$_4$ in the system A=MgO and B=SiO$_2$, we have $\nu_A=2$ and $\nu_B=1$. To put forsterite into the Gibbs free energy diagram, we must scale it by the reciprocal of $\nu_A+\nu_B=3$. The mole fraction composition of this solid thus becomes
\begin{equation}
x_s = \frac{\nu_A}{\nu_A+\nu_B}
\end{equation}
and the mole fraction weighted Gibbs free energy becomes:
\begin{equation}
\hatG^{\mathrm{sol}} = \frac{\mu^{0}_s}{\nu_A+\nu_B}
\end{equation}

We now search for the composition $x_l$ of the liquid such that the tie line connecting $x_l$ with $x_s$ (the solid composition) is the tangent to the Gibbs free energy function for the liquid (Eq.~\ref{eq-gibbs-liquid-afo-x}):
\begin{equation}\label{eq-tangent-equ}
  \left.\frac{\partial \hatG^{\mathrm{liq}}(x) }{\partial x}\right|_{x=x_l} =
  \frac{\hatG^{\mathrm{sol}}-\hatG^{\mathrm{liq}}(x_l)}{x_s-x_l}
\end{equation}
Working out the left hand side using Eq.~(\ref{eq-gibbs-liquid-afo-x}), and making use of the Gibbs-Duhem relation (cf.~equation 12 of \citet{1984GeCoA..48..661B}), we obtain:
\begin{equation}\label{eq-gradient-gibbs-act}
  \begin{split}
  \left.\frac{\partial \hatG^{\mathrm{liq}}(x) }{\partial x}\right|_{x=x_l} = &
  \left[\mu_{A}^{0,\mathrm{liq}}+RT\ln(a_A^{\mathrm{liq}})\right] \\
  &- \left[\mu_{B}^{0,\mathrm{liq}} + RT\ln(a_B^{\mathrm{liq}})\right]
  \end{split}
\end{equation}
Here we recognize the chemical potentials defined as\footnote{The interested reader will note the subtle distinction between $\partial\hatG/\partial x$, which is the derivative of the molar Gibbs free energy under the restriction that $\sum_i x_i=1$, and $\partial G/\partial n_i$, which is the partial derivative of the total Gibbs free energy with respect to the total amount of moles of component $i$.}
\begin{equation}
\mu_{A,B}^{\mathrm{liq}}(x_l) \equiv \frac{\partial G^{\mathrm{liq}}(n_{A,B})}{\partial n_{A,B}} = \mu_{A,B}^{0,\mathrm{liq}} + RT\ln(a_{A,B}^{\mathrm{liq}})
\end{equation}
meaning that we can rewrite Eq.~(\ref{eq-gradient-gibbs-act}) as
\begin{equation}\label{eq-gradient-gibbs-mu}
  \left.\frac{\partial \hatG^{\mathrm{liq}}(x) }{\partial x}\right|_{x=x_l} = \mu_A^{\mathrm{liq}}(x_l)-\mu_B^{\mathrm{liq}}(x_l)
\end{equation}
and Eq.~(\ref{eq-gibbs-liquid-afo-xab-act}) as
\begin{equation}\label{eq-gibbs-liquid-afo-muab}
  \hatG^{\mathrm{liq}}(x) = x\mu_{A}^{\mathrm{liq}}(x) + (1-x)\mu_{B}^{\mathrm{liq}}(x)
\end{equation}
Multiplying Eq.~(\ref{eq-tangent-equ}) by $(x_s-x_l)$, inserting Eq.~(\ref{eq-gradient-gibbs-mu}), and using Eq.~(\ref{eq-gibbs-liquid-afo-muab}), one can see that the $x_l$ terms cancel out and one obtains
\begin{equation}
    x_s\mu_{A}^{\mathrm{liq}}(x_l) + (1-x_s)\mu_{B}^{\mathrm{liq}}(x_l) = \hatG^{\mathrm{sol}}
\end{equation}
Multiplying again by the sum of the stoichiometric coefficients $\nu_A+\nu_B$ 
we obtain
\begin{equation}
\nu_A\mu_{A}^{\mathrm{liq}}(x_l) + \nu_B\mu_{B}^{\mathrm{liq}}(x_l) = \mu^{\mathrm{solid}}
\end{equation}
which is the equilibrium formulation using the chemical potentials. The liquidus location $x_l$ is then defined as the solution to this equation. The more general version, for arbitrary number of components, is:
\begin{equation}\label{eq-equil-using-chemical-potentials}
\sum_i\nu_i\mu_i^{\mathrm{liq}}(\mathbf{x}_l) = \mu^{\mathrm{solid}}
\end{equation}
Eq.~(\ref{eq-equil-using-chemical-potentials}) is the general formulation of the equilibrium between a fixed-composition solid phase and a liquid phase, completing the proof of the equivalence of the two formalisms.

\begin{figure}
  \includegraphics[width=1.1\linewidth]{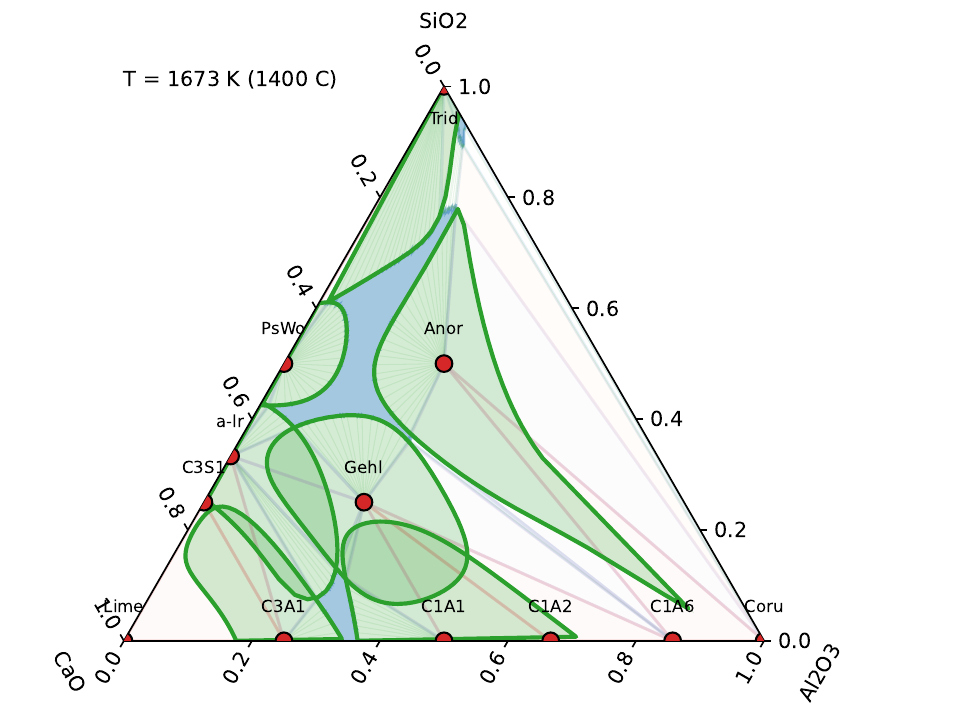}
  \caption{\label{fig-model-CaO-Al2O3-SiO2-analytic}The analytically computed liquidus curves (green lines) and solid-liquid coexistence regions (green areas) for the CaO-Al$_2$O$_3$-SiO$_2$ ternary system of Section \ref{sec-cao-al2o3-sio2}, with the results of the \phasehull{} code in the background (cf.\ Fig.~\ref{fig-model-CaO-Al2O3-SiO2-refined}). The liquidi of C1A6 and corundum are not shown, as they would dominate the figure while playing only a minor role.}
\end{figure}

To demonstrate the equivalence in practice, and at the same time test the correctness of the phase diagram created by the \phasehull{} code, Fig.~\ref{fig-model-CaO-Al2O3-SiO2-analytic} shows the liquidi of most of the stoichiometric fixed-composition phases of the 
CaO-Al$_2$O$_3$-SiO$_2$ ternary system of Section \ref{sec-cao-al2o3-sio2} at $T=1400^{\circ}$C, as computed by solving Eq.~(\ref{eq-equil-using-chemical-potentials}). The result match the results of \phasehull{}.

\section{Gibbs Minimization method}
\label{sec-app-gibbs-minimization}

Complementary to the convex hull method for computing full phase diagrams, the \phasehull{} package also offers the more commonly used Gibbs minimization method for solving, for a given bulk composition ${\mathbf x}$, the phase composition ${\mathbf Y}$ at a given temperature $T$ and pressure $P$. This method works for one single bulk composition at a time, and returns a single phase composition vector. While not ideal for producing full phase diagrams (for this, we recommend the convex hull algorithm), it is generally the more convenient method for use in simulation software. In simulations, one is typically interested in a system with a given (single) bulk composition, and how its phases evolve as temperature and pressure change. Computing full phase diagrams with dozens or thousands of phase points ${\mathbf x}_{k}$, would then be overkill. Using a classical Gibbs minimization technique for a single bulk composition ${\mathbf x}$ may then be more efficient. Furthermore, such a method does not have the scalability problem of the convex hull method when including a large number of system components (i.e., many chemical elements). It is also able to include an arbitrary number of gas-phase species in contact with the liquids and/or solids, whereas the convex hull method limits this to the number of system components.

One disadvantage of a classic Gibbs free energy minimizer, at least the one implemented in the \phasehull{} package, is that it is not easily able to properly handle miscibility gaps of liquid or solid solutions. There may be ways around it, such as implementing multiple copies of each solution into the phase composition vector, but this would be left up to the user to experiment with.

Our version of a Gibbs minimizer is implemented as follows. Consider an $M$-component system, with $N_{\mathrm{fc}}$ solid fixed-composition phases, {and/or $S$ non-ideally mixed phases (solutions), where solution $j$ consists of $M_j$ (phase-)components (for the melt phase $M_j=M$), and/or a vapor phase (or other ideal solution phase) consisting of $K$ ideally mixed molecular/atomic species. Let us define $S_M=\sum_j M_j$. We define the phase vector ${\mathbf Y}=(Y_0,\cdots,Y_{N-1})$ with $N=N_{\mathrm{fc}}+S_M+K$.} The values of ${\bf Y}$ are defined as mole fractions of each particular phase compared to the total. Therefore, the ${\bf Y}$ vector is normalized as
\begin{equation}\label{eq-gibbsmin-Ysum}
\sum_{k=0}^{N-1} Y_k = 1
\end{equation}
As always in the \phasehull{} software package, the phases corresponding to the $Y_k$ are the scaled versions of these phases (see Section \ref{sec-scaled-formula-units}), e.g., in the system of SiO$_2$ and MgO, the scaled version of the mineral Mg$_2$SiO$_3$ is Mg$_{2/3}$Si$_{1/3}$O$_{4/3}$. The values of ${\bf Y}$ can therefore be interpreted as the mole fraction of the original 1 mole of system components that is in that particular phase.

The bulk system component vector ${\mathbf x}=(x_0,\cdots,x_{M-1})$ determines the bulk composition of all coexisting phases. As usual, also this vector adds up to unity $\sum_i x_i=1$. This ${\mathbf x}$ vector is the input parameter of the Gibbs minimization scheme. The phase vector ${\mathbf Y}$ is the output.

The method we outline here is very similar to the one described extensively in the paper by \cite{2023A&A...676A..52T}. So we refer for details to that paper, and remain relatively brief here. In their work, they include stoichiometric (hence fixed-composition) solid phases and an ideal gaseous vapor phase. In our case, we additionally have a {possible} liquid phase {and possible solid solution phases}. If the liquid is ideal, {and without solid solutions,} then there is (almost) no difference. But with a non-ideal liquid phase, {and/or solid solutions,} the additional interaction term $\hatG_{\mathrm{ex}}$ in the Gibbs free energy has to be accounted for.

For the sake of simplicity, let us focus first on the case without a vapor phase, but with a liquid phase and $N_{\mathrm{fc}}>0$ fixed-composition solid phases.

For each of the fixed-composition phases, $0\le k<N_{\mathrm{fc}}$, we know its position ${\mathbf x}^{\mathrm{fc}}_{k}$ in the phase diagram, i.e., we know the mole fractions of its  components $x^{\mathrm{fc}}_{k,i}$ from the chemical formula. As usual, we normalize $\sum_{i=0}^{M-1}x^{\mathrm{fc}}_{k,i}=1$. The mole fractions of the components of the liquid ${\mathbf x}^{\mathrm{liq}}$ are part of the phase vector ${\mathbf Y}$:
\begin{equation}
x^{\mathrm{liq}}_i = \frac{Y_k}{Y_{\mathrm{liq}}},\qquad \hbox{for}\quad k=i+N_{\mathrm{fc}},\quad\hbox{and}\quad 0\le i<M
\end{equation}
where $\sum_{i=0}^{M-1}x^{\mathrm{liq}}_i=1$ and
\begin{equation}
Y_{\mathrm{liq}} = \sum_{k=N_{\mathrm{fc}}}^{N_{\mathrm{fc}}+M-1} Y_k
\end{equation}
so that $i=0,\cdots,M-1$ denote the components of the liquid. 
The total molar Gibbs free energy $\hatG$ is then the sum of the Gibbs free energies of the coexisting phases
\begin{equation}
\hatG = \sum_{k=0}^{N_{\mathrm{fc}}-1} Y_k \mu^{\mathrm{fc},0}_k + Y_{\mathrm{liq}} \hatG_{\mathrm{liq}}
\end{equation}
where $\mu^{\mathrm{fc},0}_k$ are the Gibbs free energies of the fixed-composition phases per mole of the  system components, and 
\begin{equation}
\hatG_{\mathrm{liq}} = \sum_{i=0}^{M-1} x^{\mathrm{liq}}_i \mu^{\mathrm{liq},0}_i + RT\sum_{i=0}^{M-1} x^{\mathrm{liq}}_i \ln(x^{\mathrm{liq}}_i) + \hatG_{\mathrm{ex}}({\bf x}^{\mathrm{liq}})
\end{equation}

The minimization of the Gibbs free energy must be considered under the compositional constraints that
\begin{equation}\label{eq-compositional-constraint}
\sum_{k=0}^{N_{\mathrm{fc}}-1} Y_k x^{\mathrm{fc}}_{k,i} + Y_{\mathrm{liq}}x^{\mathrm{liq}}_i = x_i
\end{equation}
for all $0\le i<M$. Given that we already impose the $\sum_k Y_k=1$ constraint, Eq.~(\ref{eq-compositional-constraint}) need only to be imposed for $0\le i<M-1$. In total, $M$ constraint equations are thus imposed. Furthermore, the values of $Y_k$ must obey $Y_k\ge 0$, which are boundary conditions we impose as well. The {\small\tt trust-constr} method of minimization {implemented in SciPy} allows both constraints and boundary conditions to be imposed.

Finally, the {\small\tt trust-constr} method requires the gradient $\partial G/\partial Y_k$ and the Hessian $\partial G/\partial Y_k\partial Y_l$ to be provided as functions. Here the partial derivatives are taken out of the plane given by $\sum_kY_k=1$, and so we use the full extensive Gibbs free energy $G$ instead of the intensive molar Gibbs free energy $\hatG$ (that is, $G(2{\mathbf Y})=2G({\mathbf Y})$ and $\hatG(\mathbf Y)=G(\mathbf Y/|\mathbf Y|)$, where $|\mathbf Y|=\sum_i^{N-1}Y_i$). The gradient (often called the Jacobian) $\partial G/\partial Y_k$ is, by definition, the chemical potential $\mu_k$ of that phase. For the fixed-composition phases it is simply $\mu^{\mathrm{fc},0}_k$. For the liquid or solid solution phase, it involves the activity of that phase in the solution. It is advantageous to have analytical functions for these activities, as any numerical errors in the gradient can cause the minimizer to become unstable. The Hessian, on the other hand, is computed, in our implementation, by numerical differentiation of the (analytical) gradient functions. 

It has turned out to be advantageous to optimize in an offset variable $Y^{\mathrm{offset}}_k=Y_k+1$ instead of $Y_k$, and in the Gibbs function internally convert back from $Y^{\mathrm{offset}}_k$ to $Y_k$. This is because most of the values of $Y_k$ become zero at the solution. The {\small\tt trust-constr} method implementation of {\small\tt scipy.optimize.minimize} seems to use relative steps compared to $|Y_k|$, which then causes it to have trouble finding the solution. With the offset, this problem disappears.

Addition of a gaseous phase is implemented the Gibbs minimizer of the \phasehull{} package in a way very similar to \citet{2023A&A...676A..52T}, except that, as we consistently do in \phasehull{}, we always use phases scaled to one mole per  system components. Also, rather than expressing the gas species in terms of their partial pressures, we consistently use molar phase fractions $Y_k$ with $k\ge N_{\mathrm{fc}}+S_M$, as for the other (solid and liquid) phases. As in the \citet{2023A&A...676A..52T} paper, the solver allows an arbitrary number of gaseous molecular species, as long as they behave as an ideal gas. The total amount of gas is then
\begin{equation}
Y_{\mathrm{gas}} = \sum_{k=N_{\mathrm{fc}}+S_M}^{N_{\mathrm{fc}}+S_M+K} Y_k
\end{equation}
where $K$ is the number of gaseous species. The partial pressures can then be retrieved as
\begin{equation}
\frac{P_k}{P} = \frac{Y_k}{Y_{\mathrm{gas}}}
\end{equation}
for $N_{\mathrm{fc}}+S_M\le k<N_{\mathrm{fc}}+S_M+K$ (the $k$ indices of the gas species), under the condition that $Y_{\mathrm{gas}}>0$ in the first place. Also, in contrast to \citet{2023A&A...676A..52T}, the $\mu^0_k$ values of the gaseous species are, in our case, computed for the pressure $P$ instead of a standard pressure. Therefore, in our method we do not need an additional pressure term in the Gibbs free energy of the gaseous species. All these differences are purely implementation choices. Mathematically both implementations are the same.

\end{appendix}

\end{document}